\documentclass[a4paper,12pt]{article}

\usepackage{amsmath,amssymb,amsfonts}
\usepackage[dvips]{graphicx}
\usepackage{epsfig}
\usepackage{color}
\usepackage{verbatim}
\usepackage{calc}
\usepackage{bbm}
\usepackage{setspace}
\usepackage{array}
\usepackage{epstopdf}
\usepackage[caption=false]{subfig}
\usepackage[left=2.4cm,top=3.3cm,right=2.4cm,bottom=3.3cm,bindingoffset=0cm]{geometry}

\newcommand{\beq}{\begin{eqnarray}}
\newcommand{\eeq}{\end{eqnarray}}
\newcommand{\bea}{\begin{eqnarray}}
\newcommand{\eea}{\end{eqnarray}}
\newcommand{\be}{\begin{equation}}
\newcommand{\ee}{\end{equation}}

\def\de{\partial}

\def\1{\mathbbm{1}}

\def\nc{\sigma}

\def\nn{\nonumber  \\}

\numberwithin{equation}{section}

\begin{document}

\title{
\begin{flushright}\ \vskip -1.5cm {\small {IFUP-TH-2018}}\end{flushright}
\vskip 20pt
\bf{ \Large Large-$N$   $\mathbb{CP}^{N-1}$  sigma model on a finite interval:  \\
general Dirichlet boundary conditions}
}
\vskip 10pt  
\author{  Stefano Bolognesi$^{(1,2)}$, 
Sven Bjarke Gudnason$^{(3)}$,\\  Kenichi Konishi$^{(1,2)}$, Keisuke Ohashi$^{(4)}$    \\[15pt]
{\em \footnotesize
$^{(1)}$Department of Physics ``E. Fermi'', University of Pisa}\\[-5pt]
{\em \footnotesize
Largo Pontecorvo, 3, Ed. C, 56127 Pisa, Italy}\\[3pt]
{\em \footnotesize
$^{(2)}$INFN, Sezione di Pisa,    
Largo Pontecorvo, 3, Ed. C, 56127 Pisa, Italy}\\[3pt]
{\em \footnotesize
$^{(3)}$Institute of Modern Physics, Chinese Academy of Sciences, Lanzhou 730000, China}\\[3pt]
{\em \footnotesize
$^{(4)}$ Research and Education Center for Natural Sciences,  Keio University}  \\[-5pt]
{\em \footnotesize  Hiyoshi 4-1-1, Yokohama, Kanagawa 223-8521, Japan }
 \\ [5pt] 
{ \footnotesize  stefanobolo@gmail.com, }
{ \footnotesize kenichi.konishi@unipi.it,} 
{ \footnotesize    bjarke@impcas.ac.cn,}  
{ \footnotesize   keisuke084@gmail.com}
}
\date{February  2018}

\maketitle

\begin{abstract}

This is the third of the series of articles on the large-$N$ two-dimensional $\mathbb{CP}^{N-1}$ sigma model, defined on a finite space interval $L$ with Dirichlet boundary conditions. Here  the cases of the general Dirichlet boundary conditions are studied, where the relative  $\mathbb{CP}^{N-1}$ orientations at the two boundaries are generic, 
 and numerical solutions are presented. Distinctive features of the
 $\mathbb{CP}^{N-1}$ sigma model, as compared e.g., to an $O(N)$
 model, which were not entirely evident in the basic properties
 studied in the first two articles in the large $N$ limit,  manifest
 themselves here.  It is found that the total energy  is minimized when the 
fields are aligned in the same direction at the two boundaries.

\end{abstract}

\newpage

\section{Introduction}

The two dimensional  $\mathbb{CP}^{N-1}$ sigma model has enjoyed a special and constant attention of  theoretical physicists 
since the pioneering work by  D'Adda et.~al.~\cite{DAdda:1978vbw}  and by Witten \cite{Witten:1978bc}.
The model is interesting as an analogous model for nonperturbative dynamics of QCD, possessing asymptotic freedom and confinement; 
it can also be related to some phenomena in condensed matter physics
such as quantum Hall effects \cite{Affleck:1984ar,Sondhi:1993zz,Ezawa:1999,Arovas:1999,Rajaraman:2002}.  

A context in which this model emerges as an effective action is the quantum excitations of the monopole-vortex composite solitons \cite{MVComplex1,MVComplex2,MVComplex3,MVComplex4}. Such systems occur under hierarchically broken gauge symmetry, e.g.,   $SU(N+1) \to SU(N) \times U(1) \to {\bf 1}$ in a color-flavor locked $SU(N)$ symmetric vacuum. The  $2D$ $\mathbb{CP}^{N-1}$ model emerges as an effective theory describing  the 
quantum fluctuations  of the collective orientational modes of the nonAbelian vortex string \cite{Hanany:2003hp,Auzzi:2003fs,Shifman:2004dr}. The vortex boundaries are the massive magnetic monopoles which are generated in the higher-mass-scale gauge-symmetry breaking, and carrying the same orientational $\mathbb{CP}^{N-1}$ moduli. In other words, magnetic monopoles are ``confined'' by the nonAbelian vortex string.  The nonAbelian monopoles arising this way are not plagued by the well-known difficulties (the topological obstruction and non-normalizable gauge zeromodes \footnote{For instance, the non-normalizable $3D$ gauge modes get converted into normalizable $2D$ modes propagating along the vortex string.}), albeit in a confinement phase.    

The presence of the boundaries implies that one is dealing with a $\mathbb{CP}^{N-1}$ model on a finite-width worldsheet, with either  Dirichlet,  Neumann 
or mixed boundary conditions, depending on the details, such as the mass hierarchy ratios.  These were part of the motivations for our previous work \cite{BKO,BBGKO}.

   More generally this type of models are interesting on their own as
   a prototype of a quantum system of mixed dimensions. 
   Since it possesses a dimensionless parameter $L \Lambda$ consisting of a length of the string $L$ and the mass gap $\Lambda$ for 
   an infinite length string,  it interpolates between  the known $\mathbb{CP}^{N-1}$ model in $2D$ in the $L\Lambda \to \infty$ limit and a quantum mechanical (classical in the case of Dirichlet model) system
   in the $L\Lambda \to 0 $ limit.  

In Refs.~\cite{BKO,BBGKO}, the generalized gap equations have been
studied analytically and solved numerically, for a wide range of
values of $L  \Lambda$.  The energy density has then been studied
carefully, by subtracting a quadratic divergence (an analogue of the
QCD vacuum energy) consistently. The results show how the dynamical
mass generation  (well known in the $2D$ $\mathbb{CP}^{N-1}$ sigma
model) and the classical $L\to 0$ limit are consistently described by
our solutions.
  In particular, it was found that the system possessed
a unique phase for any $L$, which is smoothly connected to the
``confining'' phase of the $2D$ $\mathbb{CP}^{N-1}$ model in the
$L\to\infty$ limit.
  Also, the approach to the $L\to \infty$ limit has been studied and
shown to be purely exponential, with no L\"uscher-like power-behaved
terms present.
  Finally,  the Casimir force has been studied and the presence of
different regimes (repulsive force at moderate values of
$L \sim \mathcal{O}(\Lambda^{-1})$ and attractive force at large $L$,
corresponding to a constant string tension) was predicted.

In the present work, we turn our attention to the cases in which the
relative orientation, characterized by an angle $\alpha$, of the
$\mathbb{CP}^{N-1}$ fields at the two Dirichlet  boundaries is generic
($\alpha\ne0$ in general).  The richer structure of the gap equations
is illustrated and the possible dependence of the solution on $\alpha$
is discussed. 
  These equations are then solved numerically for various values of
$\alpha$ and for different lengths $L$ of the string, and the
dependence of the total energy on the relative orientation is
investigated, both analytically and numerically.  
  We find that the total energy is minimized when the two orientations
are parallel. 
  Distinctive features of the $\mathbb{CP}^{N-1}$ sigma model manifest
themselves here,  in contrast to the first two articles, where in the
large $N$ limit the distinction from other sigma models such as the
$O(N)$ model, was not always evident, as far as the static properties
of the system were concerned.

The paper is organized as follows.
In Sec.~\ref{sec:generalBC} we generalize the case of generic
Dirichlet boundary conditions and show that the parameter space is
governed by an angle $\alpha\in[0,\pi/2]$.
In Sec.~\ref{sec:solutions} we present the new numerical method that
we have employed and the numerical solutions found.
Sec.~\ref{sec:alpha} addresses the $\alpha$-dependence of the total
energy and 
 Sec.~\ref{sec:discussion} concludes the paper with a
discussion.
Some details of our calculations have been delegated to the Appendices
\ref{sec:Care}-\ref{sec:L1and8}.

\section{General Dirichlet boundary conditions}\label{sec:generalBC}

The generalized gap equation discussed in Refs.~\cite{BKO,BBGKO} has
the form 
\beq
\frac{N}{2} \, \sum_n\frac{f_n(x)^2}{\omega_n} e^{-\epsilon \omega_n}  + \nc(x)^2 - r_{\epsilon} = 0  \,,\qquad  \partial_x^2 \nc(x) - \lambda(x) \nc(x) = 0  \,,  
\label{gapeqbb}
\eea
where
\be      r_{\epsilon}  = r_{\epsilon}^0 
 +\frac{N}{2\pi} =\frac{N}{2\pi}\left(\log\left(\frac{2}{\Lambda  \epsilon}\right)
-\gamma \right)  \;,     \label{anomaly}
\ee
and 
$r_\epsilon^0  \equiv  4 \pi/ g_{\epsilon}^2 $ stands for the bare coupling constant.
The $\mathbb{CP}^{N-1}$ fields are split into $\sigma(x)\equiv n_1(x)$
and $n_i(x)$ for $i>1$ and the latter are integrated out giving rise
to the modes $\{f_n\}$, and positive-definite energies (eigenvalues) $\{\omega_n^2| \omega_n \in \mathbb R_{>0}\}$.
The shift of $\tfrac{N}{2\pi} $  in Eq.~\eqref{anomaly}  arises from
an anomalous $\lambda$ variation  associated with the divergences in
the sum over modes \cite{BBGKO}.

The equations \eqref{gapeqbb} correspond to the Dirichlet boundary
condition
\beq
\hbox{D-D}: \qquad
n_1\!\left(-\tfrac{L}{2}\right)=n_1\!\left(\tfrac{L}{2}\right) =
\sqrt{r}\;,   \qquad
n_{i}\!\left(-\tfrac{L}{2}\right)=  n_{i}\!\left(\tfrac{L}{2}\right) =0\;,  \quad i>1\;, \label{DDbc}
\eeq
namely the orientation in the $\mathbb{CP}^{N-1}$ space was taken
to be the same at the two boundaries. 
The fields are thus defined on the finite-width worldstrip
$x\in[-\frac{L}{2},\frac{L}{2}]$; the length of the string is $L$.
The other parameter in the model is the internal scale $\Lambda$ which
also sets the energy units.
The only physical parameter is thus $L\Lambda$. 

In the present paper,  the cases in which the fields are orientated differently in the $\mathbb{CP}^{N-1}$ space  at the two boundaries   will be studied.
Due to the global $SU(N)$ symmetry of the  $\mathbb{CP}^{N-1}$ model and by the definition of the $\mathbb{CP}^{N-1}$ coordinates
\be   
n_i(x) \sim e^{i \theta (x)} n_i(x) \;,  \label{U1sym}
\ee
the most general D-D type  boundary condition can be taken in the form,
\beq   
\left(\begin{array}{c}n_1\!\left(\tfrac{L}{2}\right) \\n_2\!\left(\tfrac{L}{2}\right)\end{array}\right) =  \left(\begin{array}{c} 1 \\0\end{array}\right) \sqrt {r_{\epsilon}} \;;
\label{bc1} 
\eeq 
\beq 
  \left(\begin{array}{c}n_1\!\left(-\tfrac{L}{2}\right) \\n_2\!\left(-\tfrac{L}{2}\right)\end{array}\right) = \left(\begin{array}{cc}e^{i \gamma} \cos \alpha  & e^{i \beta} \sin \alpha \\- e^{-i\beta}\sin \alpha & e^{-i \gamma} \cos \alpha \end{array}\right) \left(\begin{array}{c} {\sqrt{r_{\epsilon}}}  \\0\end{array}\right)  \sim 
   \left(\begin{array}{c} \cos \alpha  \\   \sin \alpha\end{array}\right)  \sqrt {r_{\epsilon}}\;;    \label{bc2}
\eeq
\beq   
n_{i}\!\left(-\tfrac{L}{2}\right)=  n_{i}\!\left(\tfrac{L}{2}\right)=0\;,  \qquad i>2\;.    \label{bc2bis} 
\eeq
Accordingly, the generalized gap equation becomes
\beq
\frac{N}{2} \, \sum_n\frac{f_n(x)^2}{\omega_n} e^{-\epsilon \omega_n}  + |\sigma_1(x)|^2 +   |\sigma_2(x)|^2   - r_{\epsilon} = 0  \,,  \label{gapeqbbGen1} \eeq
\beq
 \partial_x^2 \sigma_1 - \lambda(x) \sigma_1 = 0  \,,  \qquad   \partial_x^2 \sigma_2 - \lambda(x) \sigma_2 = 0\;,
\label{gapeqbbGen2}
\eea
together with the boundary conditions \eqref{bc1}-\eqref{bc2bis} and
we have defined the two ``classical'' fields: $\nc_1\equiv n_1$ and
$\nc_2\equiv n_2$. 
$\{f_n(x), \omega_n^2 \}$ are the solutions of the Schr\"odinger equation
\beq
\left( - \partial_{x}^2  + \lambda(x) \right) f_n(x) = \omega_n^2
\ f_n(x)   \;, \qquad  \int_{-\frac{L}{2}}^{\frac{L}{2}}   dx \,
f_n(x) f_m(x) = \delta_{n\,m}\;.
\label{operator}
\eeq

Actually it is useful to start with the observation that the differential equation
\beq  \partial_x^2 \nc(x) - \lambda(x) \nc(x) = 0 \;, \label{difeq}
\eeq
for any given $\lambda(x)$  has two linearly independent solutions.  A particular convenient choice turns out to be the solution $\sigma_R(x)$  and $\sigma_L(x) $ with the following characteristics.  Near the left boundary,  $x=- \tfrac{L}{2}$, the two solutions behave as \cite{BKO}
\beq 
\sigma_L(x) \sim     \sqrt{ \tfrac{N}{2\pi} \log \tfrac{L_0}{x+L/2}}  \;, \qquad  
 \sigma_R(x) \sim   W    \frac{x+L/2}{  \sqrt{  \tfrac{N}{2\pi} \log \tfrac{L_0}{x+L/2}} }\;,   \label{nearby1}
\eeq 
whereas near  $x= \tfrac{L}{2}$,
\beq 
\sigma_R(x) \sim    \sqrt{   \tfrac{N}{2\pi}  \log \tfrac{L_0}{L/2-x}}  \;, \qquad   \sigma_L(x) \sim   W  \frac{L/2-x}{  \sqrt{   \tfrac{N}{2\pi}  \log \tfrac{L_0}{L/2-x}} }\;.   \label{nearby2}
\eeq 
The normalization of the divergent behavior (the first term in Eqs.~\eqref{nearby1} and \eqref{nearby2}) is fixed by the gap equation,  $W$ is a constant determined later,  and $L_0$ is a certain Length parameter.\footnote{ Strictly speaking, we need to introduce another UV cutoff parameter $\epsilon_b$ to impose condition (\ref{bc1}) and condition (\ref{bc2}) at a distance $\epsilon_b$ from one of the boundaries.  Using Eq.~(\ref{anomaly}),  
we find the two UV cutoff parameters are related to each other as
\begin{eqnarray}
\frac{L_0}{\epsilon_b}= \frac{2 e^{-\gamma}}{\Lambda \epsilon}\, \left(= \frac{2\pi n_{\rm max}}{L \Lambda}\right).
\end{eqnarray}
That is,  if we fix a ratio $\epsilon_b/\epsilon$ in the limit of $\epsilon \to 0$, $L_0$ is a given parameter and thus, 
 independent of $L$ and $\alpha$.
} 
For even $\lambda(x)$,  given a solution $\sigma(x)$, the parity transformed function  $\sigma(-x)$  is also a solution. It follows that 
\beq     \sigma_L(x)=\sigma_R(-x)\;.
\eeq
Exactly at the boundaries, the gap equation tells us that 
\be       \sigma_R(\tfrac{L}{2})  =  \sqrt{r_{\epsilon}}\;,    \qquad    \sigma_R(- \tfrac{L}{2})  =  0\;,  
\ee
and similarly
\be       \sigma_L(-\tfrac{L}{2})  =  \sqrt{r_{\epsilon}}\;,    \qquad    \sigma_L(\tfrac{L}{2})  =  0\;.  
\ee
 The Wronskian of  the two solutions are defined by
\be   W   \equiv     \sigma_L(x)  \sigma_R'(x) -   \sigma_R(x)  \sigma_L'(x)   = {\rm const}  >  0\;.  \label{Wrnsk}
\ee
By evaluating the Wronskian (which is constant) near the two boundaries,    the constant in Eqs.~\eqref{nearby1} and \eqref{nearby2} is seen to be precisely  the Wronskian itself.   Let us define also  the even and odd solutions:
\beq    \sigma^{(e)}(x) \equiv  \sigma_R(x)+  \sigma_L(x)\;;  \qquad   \sigma^{(o)}(x) \equiv \sigma_R(x)-  \sigma_L(x)\;.
\label{eando}\eeq 

In terms of  $\sigma_R(x)$ and $\sigma_L(x)$,  the most general solution can be written as
\begin{eqnarray}
\sigma_a(x) =  q^{\rm R}_a \sigma_R(x)+q^{\rm L}_a \sigma_L(x)
\;,  \qquad a=1,2,\ldots, N+1\;,   \label{mostgen}
\end{eqnarray}
which corresponds to the pair of boundary conditions
\begin{eqnarray}
\sigma_a\left(\tfrac{L}2\right)= q^{\rm R}_a \sqrt{r_\epsilon}\;,\qquad  \sigma_a\left(-\tfrac{L}2\right)= q^{\rm L}_a \sqrt{r_\epsilon}\;.   \label{genecond1}
\end{eqnarray}
In other words,   the vectors  $q_a^{\rm R}$ and $q_a^{\rm L}$, 
\be \sum_a|q_a^{\rm R}|^2=\sum_a|q_a^{\rm L}|^2=1\;,   \ee
describe {\it  the orientations of the classical field at the right and left boundaries, respectively. }
By using the global $SU(N)$ invariance of $\mathbb{CP}^{N-1}$, one may choose  
\be    q^{\rm R}_a=  \left(\begin{array}{c}1 \\0 \\0 \\\vdots \\0\end{array}\right)\;, \qquad  q^{\rm L}_a= \left(\begin{array}{c}\cos \alpha \\\sin \alpha \\0 \\\vdots \\0\end{array}\right)\;,\label{genecond2}
\ee
and this  corresponds to the general boundary condition,  anticipated in Eqs.~\eqref{bc1}-\eqref{bc2bis}.

\begin{figure}
\begin{center}
\includegraphics[width=3.7  in]{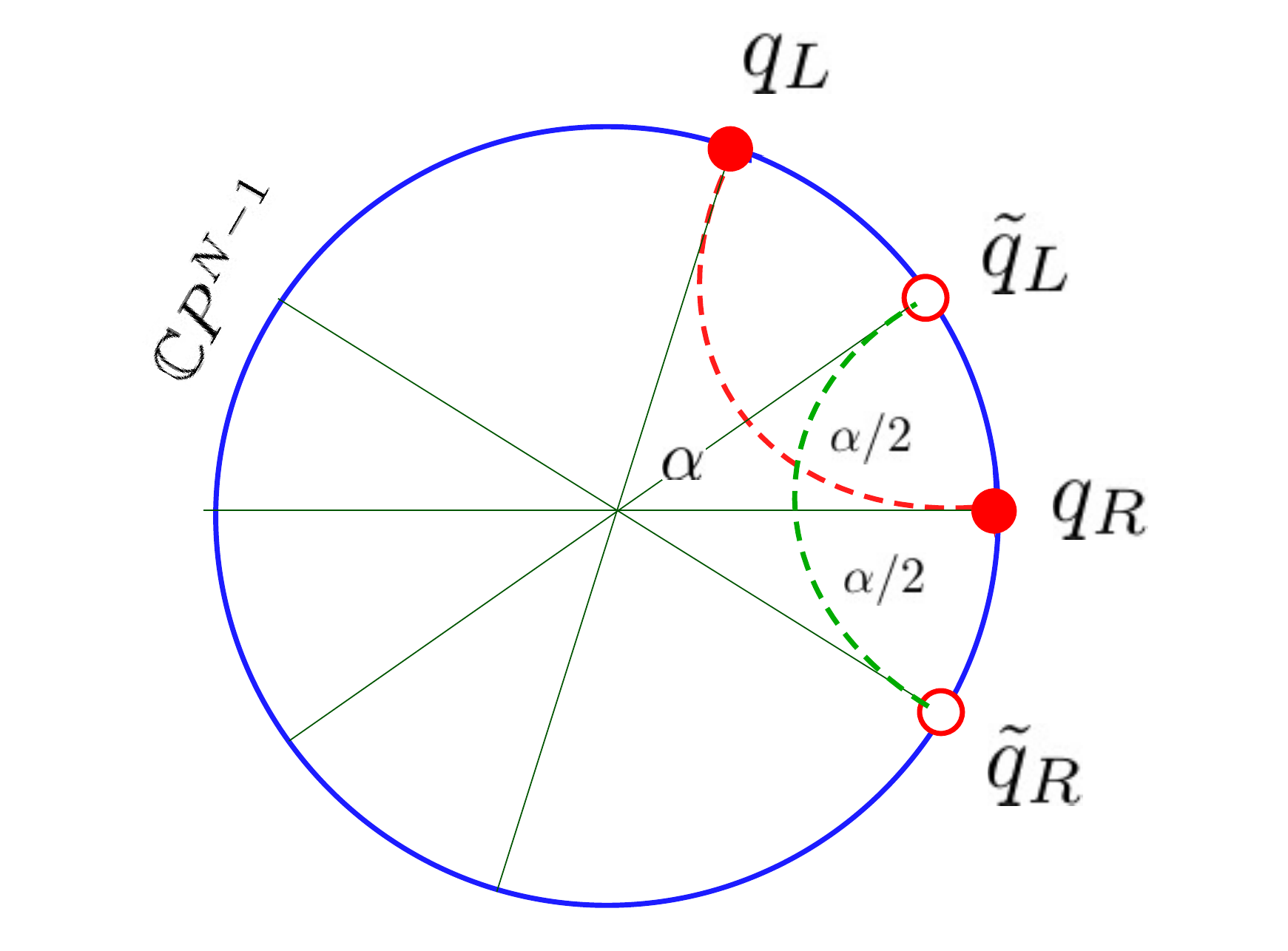}
\caption{\small  Boundary conditions  (\ref{genecond2}) and
  (\ref{genecond22}).  The orientation vectors in the two cases are
  simply related by a rotation by $\alpha/2$ of the 
$\mathbb{CP}^{N-1}$ frame and are clearly equivalent.  The dashed lines show schematically the movement of the classical field from one boundary to the other.   The circles stand for the $\mathbb{CP}^{N-1}$ points to which the fields approach at the boundaries. }
\label{New4}
\end{center}
\end{figure}

The general boundary condition may be put in a form which looks more symmetric with respect to the exchange of the two boundaries. 
 This can be done by rotating the $\mathbb{CP}^{N-1}$ frame by the angle $\alpha/2$. By writing only the first two components of Eqs.~\eqref{bc1}-\eqref{bc2bis}, 
\begin{eqnarray}
\left(\begin{array}{cc}
\sigma_1(x)\\ \sigma_2(x)
\end{array}\right)&=&
\left(\begin{array}{cc}
1& \cos \alpha \\ 0 &\sin \alpha
\end{array}\right) \left(\begin{array}{cc}
\sigma_R(x) \\ \sigma_L(x)
\end{array}\right)\nn
&=&\frac12 \left(\begin{array}{cc}
1+\cos \alpha  & 1-\cos \alpha\\ \sin \alpha &-\sin \alpha
\end{array}\right) \left(\begin{array}{cc}
\sigma_{\rm e}(x)\\ \sigma_{\rm o}(x)
\end{array}\right)\nn
&=& \left(\begin{array}{cc}
\cos \frac{\alpha}2  & -\sin \frac{\alpha}2 \\ \sin \frac{\alpha}2 & \cos \frac{\alpha}2
\end{array}\right) 
\left(\begin{array}{cc}
{\tilde \sigma}_1(x)\\ {\tilde \sigma}_2(x)
\end{array}\right)\;,   \label{generalbc}
\end{eqnarray}
where 
\be 
\left(\begin{array}{cc}
{\tilde \sigma}_1(x)\\ {\tilde \sigma}_2(x)
\end{array}\right)
=
 \left(\begin{array}{cc}
\cos \frac{\alpha}2 \,  \sigma_{\rm e}(x)\\ -\sin \frac{\alpha}2  \,  \sigma_{\rm o}(x)
\end{array}\right)\;.   \label{geneconsigmatilde} 
\ee
 ${\tilde \sigma}_1$ and ${\tilde \sigma}_2$ are the components of the classical field  
in the new  $\mathbb{CP}^{N-1}$ frame.  
By using the orientation vectors, this means that 
  \begin{eqnarray}
{\tilde \sigma_a}\left(\frac{L}2\right)={\tilde  q}^{\rm R}_a \sqrt{r_\epsilon},\qquad  {\tilde \sigma}_a\left(-\frac{L}2\right)= {\tilde q}^{\rm L}_a \sqrt{r_\epsilon}\;,   \label{genecond11}
\end{eqnarray}
with
\be    {\tilde q}^{\rm R}_a  =  \left(\begin{array}{c} \cos \frac{\alpha}{2}   \\  \sin \frac{\alpha}{2} \\0 \\\vdots \\0\end{array}\right)\;,   \qquad  {\tilde q}^{\rm L}_a= \left(\begin{array}{c}\cos \frac{\alpha}{2} 
\\ -\sin \frac{\alpha}{2}   \\0 \\\vdots \\0\end{array}\right)\;.\label{genecond22}
\ee
See Fig.~\ref{New4}.

\subsection{The gauge field \label{gauge}}

In our first two papers \cite{BKO,BBGKO}, the $A_0=0$ gauge was chosen, which is appropriate for studying static vacuum configurations.
 Also the gauge field $A_x$  was assumed to be absent in the vacuum (the functional stationary point).
Note that the equation of motion for  $A_x$ is
\bea
\frac{\delta S}{\delta A_x}
&=& \langle  J_{\mu} \rangle
= \Big\langle i\sum_i(n^{\dagger} \partial_x n_i -  \partial_x
n^{\dagger}   n_i ) \Big\rangle +   2   A_x  \, \langle
n^{\dagger} n \rangle   \nonumber\\
&=& i   (\sigma^{\dagger}\partial_x   \sigma - \partial_x
\sigma^{\dagger}\sigma) + 2 A_x r_{\epsilon} =0\;,
\eea
where $n$ and $\sigma$ are column vectors and hence $n^\dag n$ is the
inner product.
 $A_x$  then satisfies  
\be     A_x=  -\frac{i}{2 \,r_{\epsilon}} (\sigma^{\dagger}\partial_x   \sigma  - \partial_x \sigma^{\dagger}\sigma  ) \;, \label{shows}
\ee
and this vanishes  if the solution for $\sigma$
is real.    More precisely,  as equation (\ref{difeq}) is real  ($\lambda(x)$ is real),   the field  $\sigma$ can always be chosen to be real by 
the local $U(1)$ gauge transformation. Eq.~\eqref{shows} shows that $A_x$ vanishes in the vacuum in such a gauge; a fact used in Refs.~\cite{BKO,BBGKO}. 
 
 Similarly, the configuration with the most general boundary condition   (\ref{mostgen})   gives 
the following (classical part of the) $U(1)$ current
\begin{eqnarray}
J^{U(1)}_x=i\sum_a \big(\bar \sigma_a(x) \sigma_a'(x)-\bar \sigma_a'(x) \sigma_a(x)\big)=iW \sum_a (\bar q_a^{\rm R}  q_a^{\rm L}
-\bar q_a^{\rm L}  q_a^{\rm R})\;.
\end{eqnarray}
Thus if one chooses $q_a^{\rm R}, q_a^{\rm L}$ such  that   $J_x^{U(1)}$ is  non-vanishing, then 
a non-vanishing constant gauge field $A_x=-J_x^{U(1)}/2r_\epsilon \not=0$ is predicted by the EOM for $A_x$. 
After making the $U(1)$ gauge transformation so that $A_x=0$, the right hand side of the above must vanish.  Our choice with real 
 $q_a^{\rm R}, q_a^{\rm L}$, in Eq.~\eqref{genecond2}, thus yields
 $J_x^{U(1)}=0$,  corresponding to the choice of gauge $A_x=0$.

\subsection{Range of $\alpha$}\label{sec:range}

Exchanging the boundaries $x=\frac L 2$ and  $x=-\frac L 2$ and using an appropriate rotation, 
a configuration with $\alpha$ is seen to be equivalent to the one with $-\alpha$. 

Also,  two solutions with $\alpha$ and with $\pi -\alpha $ can be regarded as the same boundary conditions.  
For consider the left boundary
\be  q^{\rm L}_a= \left(\begin{array}{c}\cos \alpha \\\sin \alpha \\0 \\\vdots \\0\end{array}\right)\;.  \ee
The change   $\alpha \to \pi -\alpha $  makes
\be    q^{\rm L}_a \to   \left(\begin{array}{c}  - \cos \alpha \\ \sin \alpha \\0 \\\vdots \\0\end{array}\right) \sim
 \left(\begin{array}{c} - \cos \alpha \\ -\sin \alpha \\0 \\\vdots \\0\end{array}\right) \sim   \left(\begin{array}{c}  \cos \alpha \\  \sin \alpha \\0 \\\vdots \\0\end{array}\right)  
\ee
where we have used a global  $SU(2) \subset SU(N)$ e.g.,  in the $2-3$ plane first (which does not affect $q^{\rm R}_a$), and a $U(1)$ equivalence relation $q^{\rm L}_a \sim e^{i\beta}q^{\rm L}_a$
at the end.   

Note that in the $\mathbb{CP}^{N-1}$ sigma model the $U(1)$  symmetry (\ref{U1sym}) is implemented as a {\it  local gauge symmetry}.    Therefore the solution  with the boundary condition
\be    q^{\rm R}_a=  \left(\begin{array}{c}1 \\0 \\0 \\\vdots \\0\end{array}\right)\;, \qquad  q^{\rm L}_a= \left(\begin{array}{c} - \cos \alpha \\ -  \sin \alpha \\0 \\\vdots \\0\end{array}\right)\;,\label{genecond2Bis}
\ee
can be regarded as a gauge transformation of the solution (\ref{genecond2}). In order to have the same physics (e.g.~the same energy density, etc.), however, one must appropriately introduce 
the gauge field $A_x$. Note that even though the $\sigma_1$ field is chosen to take real values at the two boundaries, it is necessary that it goes through a phase rotation, 
\be    \sigma_1(x) = e^{i \beta(x)}  |\sigma_1(x)|\;, \qquad  \beta(\tfrac{L}{2})= 0\;, \quad  \beta(-\tfrac{L}{2})= \pi\;,
\ee
meaning that such a solution necessarily generates a current
$J_x(x)$, hence a nonvanishing gauge field $A_x$ (see the previous Subsection).   Once they are appropriately taken into account,  
the $\pi - \alpha$ solution (\ref{genecond2Bis}) is simply a gauge transformation of the $\alpha$ solution, (\ref{genecond2}). 

Accordingly, the range of the parameter $\alpha$  can be taken as   
\be  0 \le   \alpha   \le  \frac{  \pi }{2}   \label{range}    \ee
 without loss of generality.  The result of  Section~\ref{sec:alpha} indeed  indicates that the solutions with $\alpha$ in this 
 range are the stable ones.

Nevertheless, the solution  with  $\alpha$ and the one with
$\pi-\alpha$ are distinct if  $A_x\equiv 0$ and if   $\sigma_1$ and
$\sigma_2$  are kept real.  In section \ref{sec:solutions}, we will discuss
the numerical solutions of the gap equation (\ref{gapeqbbGen1}) and
\eqref{gapeqbbGen2} for the whole range of $\alpha$,  i.e.~$[0, \pi]$. 
Numerically, this is advantageous because we can avoid introducing the
gauge field $A_x$ and we can keep the $\sigma$ fields real in the
calculations.

\subsection{The ``solutions'' with $\alpha=0$ and $\alpha=\pi$}

A particular case of interest about the gauge (non-)equivalence of the solutions with $\alpha$ and $\pi - \alpha$,  concerns the solutions 
with $\alpha=0$ and $\alpha=\pi$. 
The classical field  $\sigma_a(x)$ as a function of $x \in [ -\tfrac{L}{2},  \tfrac{L}{2}] $  can be regarded as a path  from a point  in  $\mathbb{CP}^{N-1}$  to another, through the ``interior'' of  it: in the space
$\mathbb{CP}^{N-1}\times {\mathbb R}$. 
The solution $\alpha=0$  corresponds to the movement from the left to the right boundary, as
\beq   \left(\begin{array}{c}\sigma_1(x) \\ \sigma_2(x) \\ \vdots   \end{array}\right) =  \left(\begin{array}{c} 
\, \sigma^{(e)}(x)  \\  0  \\ \vdots   \end{array}\right) \;;       \qquad   \left(\begin{array}{c} 
\, 1 \\  0   \\ \vdots  \end{array}\right) r_{\epsilon}  \, \longrightarrow    \left(\begin{array}{c} 
\, 1 \\  0  \\ \vdots   \end{array}\right) r_{\epsilon}  \,  :
\label{bc4}  \eeq
this is the case studied in Refs.~\cite{BKO,BBGKO}.  It describes a closed path in  $\mathbb{CP}^{N-1}\times {\mathbb R}$.
Other solutions $0 < \alpha <  {\pi} $ involve two components $\sigma_1$ and $\sigma_2$ nontrivially, and in general do not describe a closed path.

It is interesting to consider  the solution with 
another particular  boundary condition  $\alpha=\pi$:  it  looks like  (by using the first line of Eq.~\eqref{generalbc})
\beq   \left(\begin{array}{c}\sigma_1(x) \\ \sigma_2(x) \\ \vdots  \end{array}\right) =  \left(\begin{array}{c} 
\,  \sigma^{(o)}(x)   \\   0 \\ \vdots   \end{array}\right) \;;       \qquad   \left(\begin{array}{c} 
\, -1 \\  0 \\ \vdots    \end{array}\right) r_{\epsilon}  \,  \longrightarrow    \left(\begin{array}{c} 
\,  1  \\   0  \\ \vdots   \end{array}\right) r_{\epsilon}  \,  \;.
\label{bc5}  \eeq
It may appear that the two solutions  (\ref{bc4}) and (\ref{bc5}) correspond to topologically
inequivalent classes of paths  and that  both solutions might hence be  stable.  
Actually, $(1,0,\ldots)$ and  $(-1,0,\ldots) $ represent the same point of $\mathbb{CP}^{N-1}$.\footnote{As discussed in Subsection~\ref{sec:range},  the two solutions would be simply gauge transform of each other if the gauge field is appropriately taken into account in (\ref{bc5}).    Here we are discussing the two solutions, both with $A_x\equiv 0$, hence not gauge-equivalent. 
}
 In other words  the $\alpha=\pi$ solution also describes a closed path with the same starting and end points as in the $\alpha=0$ solution.   
 Since the space $\mathbb{CP}^{N-1}\times {\mathbb R}_{>0}$ is simply connected,  these two closed loops are  homotopically equivalent.  
Therefore only one of the solutions can be stable.   The study of the $\alpha$ dependence of the energy  in Section~\ref{sec:alpha}  indicates that the solution with $\alpha=0$ is indeed the stable one.



%
%

\section{The solutions of the gap equation}\label{sec:solutions}

With the most general D-D boundary condition,   the equations to be solved are
\bea
\frac{N}{2} \, \sum_n\frac{f_n(x)^2}{\omega_n} e^{-\epsilon \omega_n}  + \nc_1(x)^2  + \nc_2(x)^2   - r_{\epsilon} = 0  \,,      \label{gapeq1}  \\
 \partial_x^2 \nc_1(x) - \lambda(x) \nc_1(x) = 0  \,,  \qquad
 \partial_x^2 \nc_2(x) - \lambda(x) \nc_2(x) = 0 \,, \label{eq:sigmaeom}
\eea
together with the boundary conditions \eqref{genecond1} and
\eqref{genecond2}. 
In Ref.~\cite{BBGKO}, we used a random-walk method for solving the
first of the above equations.
In this paper, however, we use a relaxation method akin to the
procedure of solving the time evolution in a heat-like equation.
In order to derive the equation that we will solve numerically, let us
start from the energy functional of the form \cite{BBGKO}
\bea
E = N \sum_n \omega_n e^{-\epsilon\omega_n}
+ \int_{-\frac{L}{2}}^{\frac{L}{2}} dx
\left[\sigma_1'(x)^2 + \sigma_2'(x)^2
  + \lambda\left(\sigma_1(x)^2 + \sigma_2(x)^2 - r_\epsilon^0\right)
  + \mathcal{E}_{\rm uv}\right],
\eea
from which a variation with respect to $\lambda$ yields
\bea
\frac{\delta E}{\delta\lambda} =
\frac{N}{2}\sum_n \frac{f_n(x)^2}{\omega_n} e^{-\epsilon\omega_n}
+ \sigma_1(x)^2 + \sigma_2(x)^2 - r_\epsilon \,;
\eea
indeed this is how the generalized gap equation was derived in
Ref.~\cite{BBGKO}.
Although the above equation is formally convergent due to the
$\epsilon$ regulator, it is not practical for numerical calculations
as it still includes an infinite number of modes.
Therefore, we will switch regularization to a finite number of modes,
$n\leq n_{\rm max}$.
For consistency, we need to modify the coupling $r_\epsilon$
accordingly
\bea
\frac{\delta E}{\delta\lambda} =
\frac{N}{2}\sum_{n=1}^{n_{\rm max}} \frac{f_n(x)^2}{\omega_n}
+ \sigma_1(x)^2 + \sigma_2(x)^2 - r_{n_{\rm max}} \,,
\label{eq:dEdlambda_nmax}
\eea
where
\bea
r_{n_{\rm max}}=\frac{N}{2\pi}
\log\left(\frac{2\pi n_{\rm max}}{L\Lambda}\right),
\label{eq:rnmax}
\eea
was defined in Ref.~\cite{BKO}.

Now, instead of setting Eq.~\eqref{eq:dEdlambda_nmax} equal to zero
(which it should be), we will introduce a fictitious time dependence
and flow said fictitious time evolution
\bea
\frac{\delta E}{\delta\lambda}
= \frac{\de\lambda}{\de\tau}
= \frac{\lambda_{t+1} - \lambda_{t}}{h_\tau}\,,
\eea
where in the last equality, we have introduced a discretized
first-order time derivative, $h_\tau$ is the time step, and $t$ is an
index of the current time slice.
Rearranging, we can finally write the formal evolution equation
\bea
\lambda_{t+1} = \lambda_t
+ h_\tau\left[
  \frac{N}{2}\sum_{n=1}^{n_{\rm max}} \frac{f_n(x)^2}{\omega_n}
  + \sigma_1(x)^2 + \sigma_2(x)^2 - r_{n_{\rm max}}
  \right]\,,
\label{eq:lambdaevolution}
\eea
where $h_t$ should be chosen appropriately; in particular, it should
be small enough to ensure convergence of the algorithm. 

We are now almost ready to perform the numerical calculations.
First we discretize the string into a one-dimensional lattice.
We need to calculate the Schr\"odinger modes and energies,
$\{f_n,\omega_n^2\}$ from Eq.~\eqref{operator}, and we will do that
simply by diagonalizing the discretized version of the operator
$(-\de_x^2+\lambda)$ subject to the boundary conditions
$f_n(\pm\tfrac{L}{2})=0$.
Next, we will solve the equations of motion for $\sigma_{1,2}$,
\eqref{eq:sigmaeom}, by back-solving the discretized version of the
operator $(-\de_x^2+\lambda)$ on their respective boundary conditions
(see Ref.~\cite{BBGKO} for details);
in particular, we will use the conditions of
Eqs.~\eqref{genecond1} and \eqref{genecond2}. 
We start the algorithm by using a guess for $\lambda$.
The final step is to calculate the new $\lambda$ from
Eq.~\eqref{eq:lambdaevolution}.
This is the end of the cycle; now we start over by calculating the
modes and energies and so on. 
The cycle continues until the integral of the absolute value
of Eq.~\eqref{eq:dEdlambda_nmax} is smaller than an appropriate small 
number, which we shall call $\varepsilon_{\rm numerical}$, see
Appendix \ref{app:numacc}.  

\begin{figure}[!ht]
\begin{center}
\mbox{\subfloat{\includegraphics[width=0.49\linewidth]{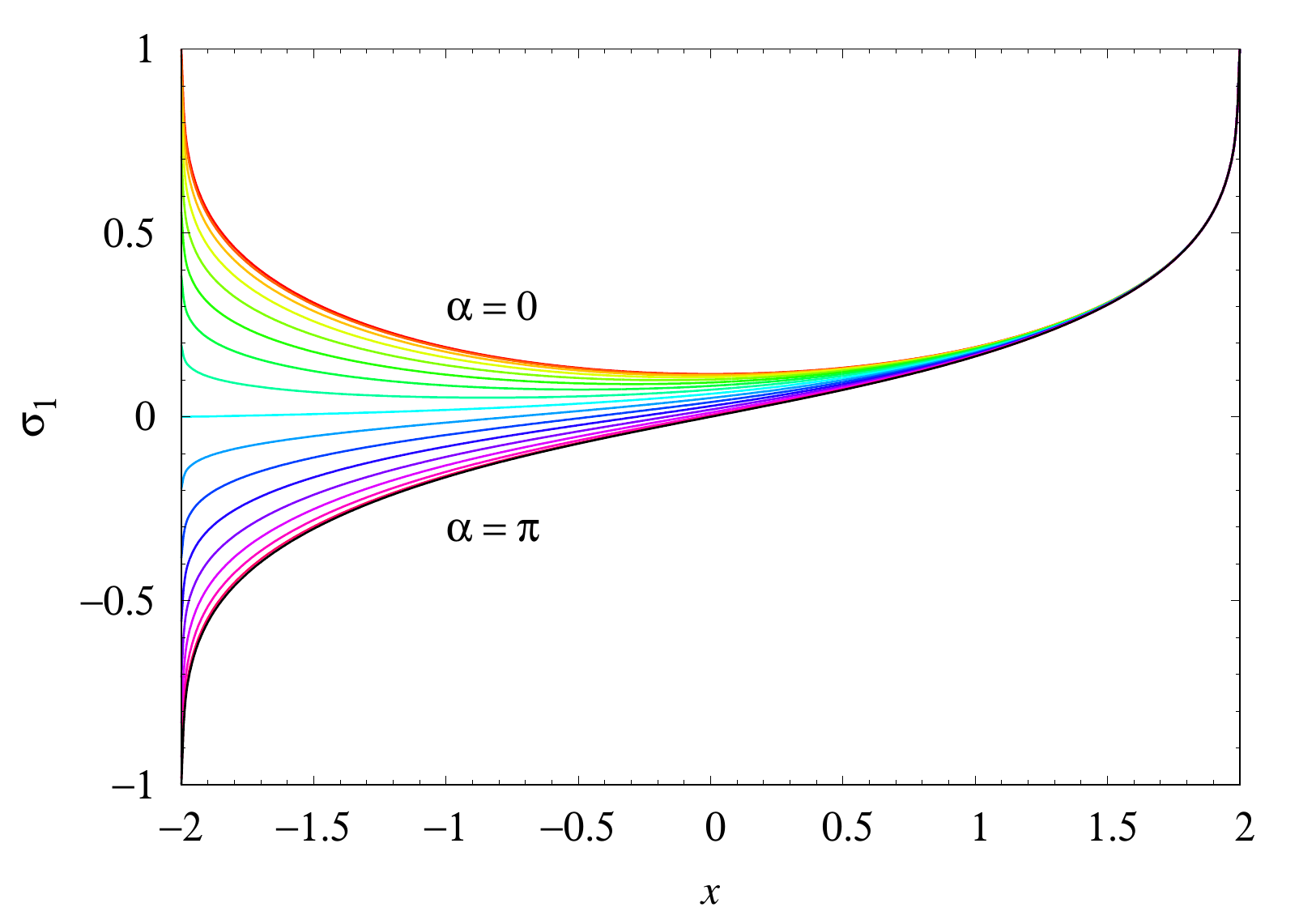}}
\subfloat{\includegraphics[width=0.49\linewidth]{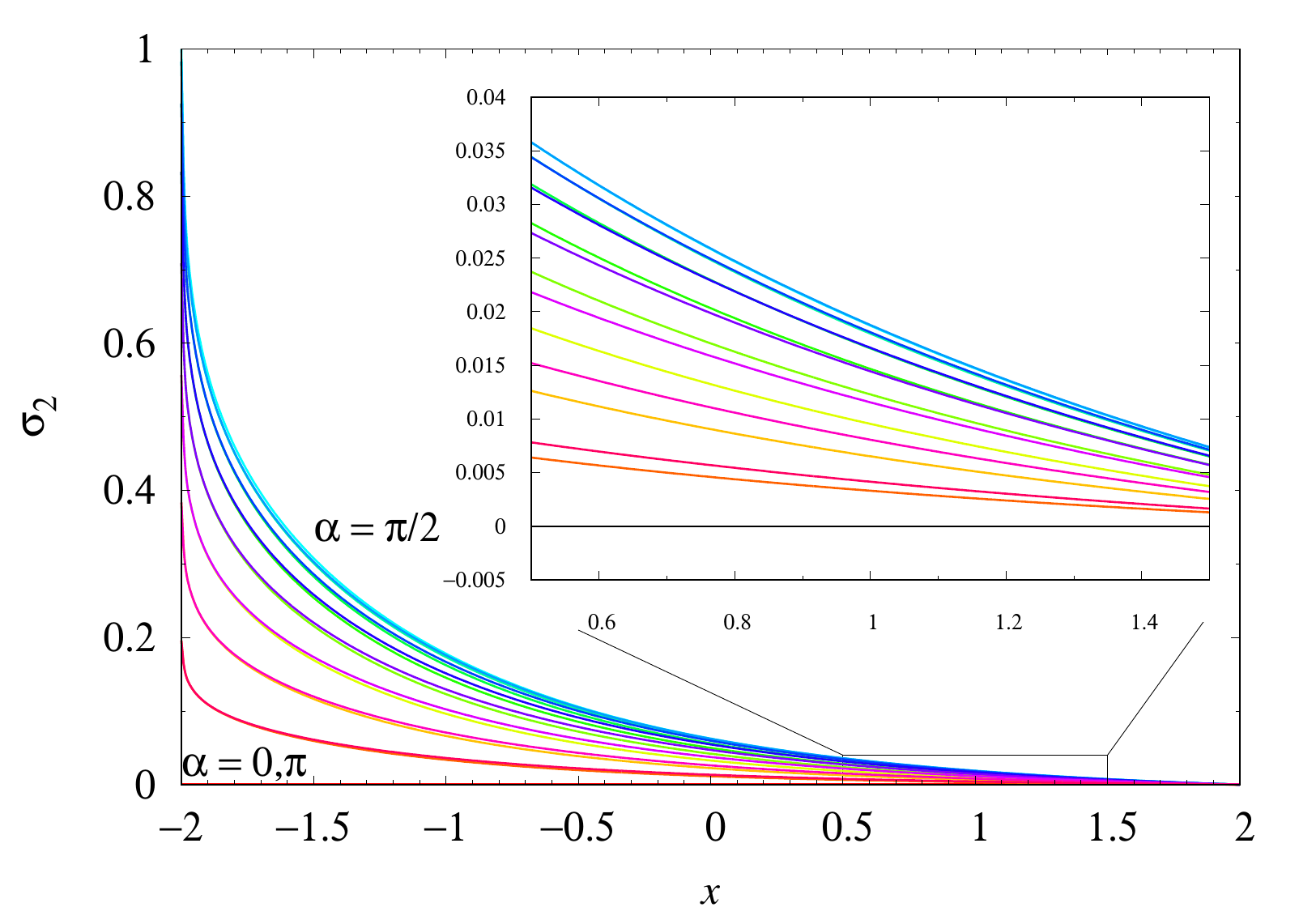}}}
\caption{\small  The functions  $\sigma_1(x)$ (left) and
  $\sigma_2(x)$ (right)  which are the solutions to the gap equation,
  Eq.~(\ref{gapeq1}), for various values of $\alpha$ for  $L=4$ and
  $\Lambda=1$.  $\alpha=0, \tfrac{\pi}{16}, \tfrac{2 \pi}{16},\ldots,\pi$
  from the top to the bottom curves for $\sigma_1$ on the left. 
  The curves for $\sigma_2(x)$ on the right read from the bottom
  ($\sigma_2\equiv 0$  for $\alpha=0$) to the top curve at
  $\alpha=\pi/2$, and then back towards the bottom ($\sigma_2\equiv0$
  again for $\alpha=\pi$).
  The colors are shown in the legend of Fig.~\ref{fig:sigmatilde}.
}
\label{fig:sigmas}
\end{center}
\end{figure}

\begin{figure}[!ht]
\begin{center}
\mbox{\subfloat{\includegraphics[width=0.49\linewidth]{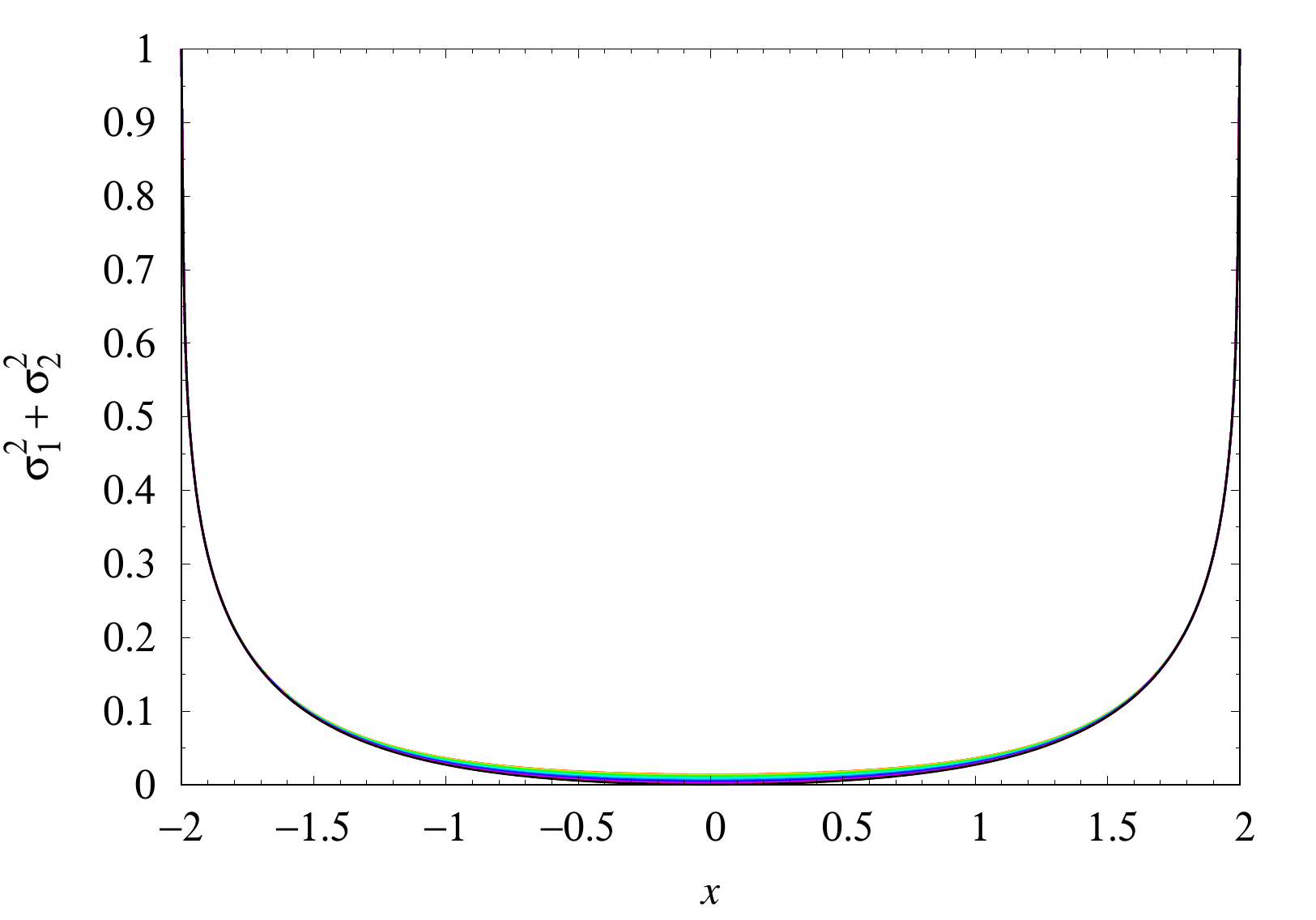}}
\subfloat{\includegraphics[width=0.49\linewidth]{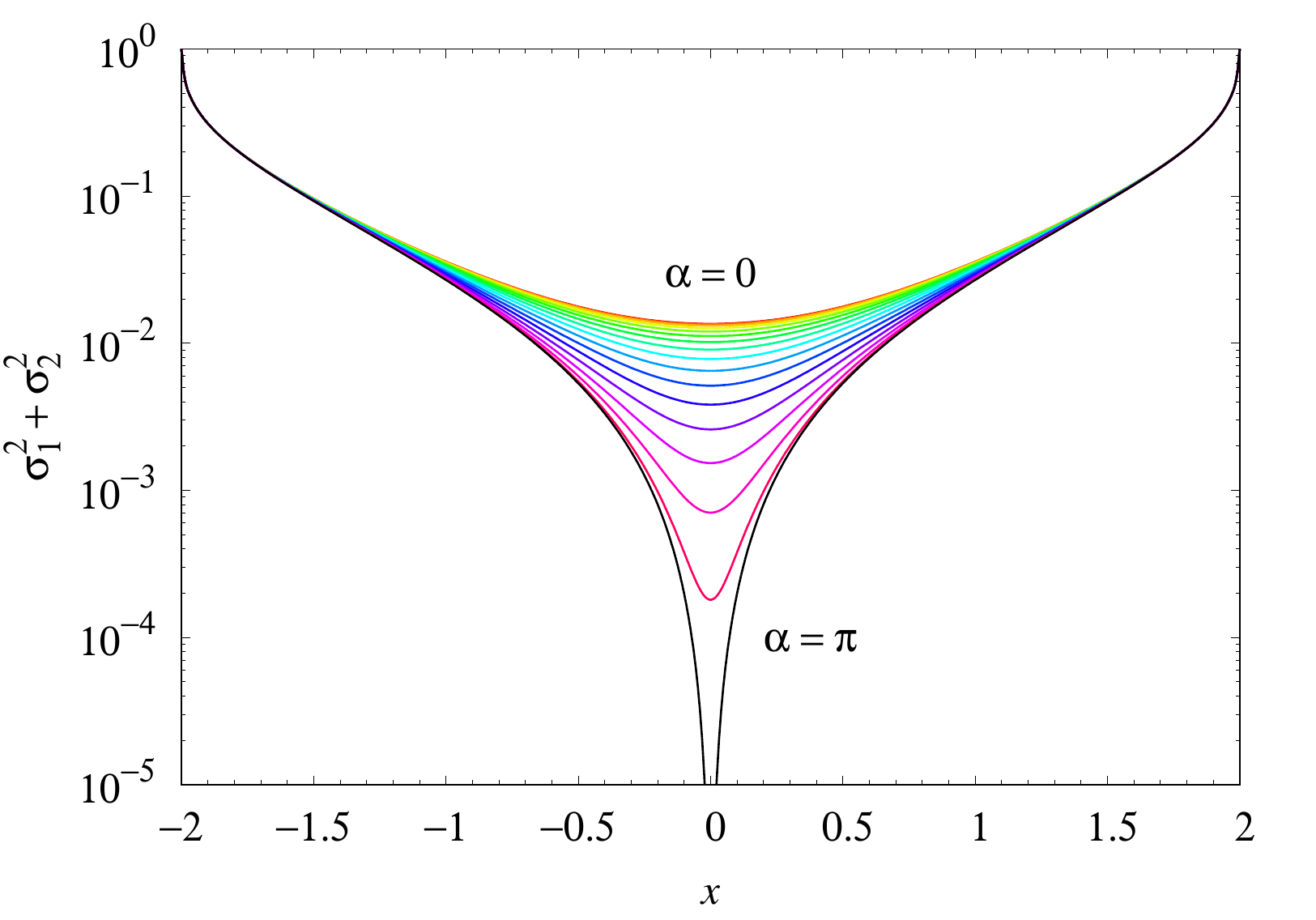}}}
\caption{\small The function  $\sigma_1(x)^2+\sigma_2(x)^2$  is shown
  in a normal plot (left) and on a logarithmic scale (right), for
  various values of $\alpha$ for  $L=4$ and  $\Lambda=1$.  The values
  of $\alpha$ are
  $0, \tfrac{\pi}{16}, \tfrac{2 \pi}{16},\ldots, \pi$ from the top to
  the bottom curves.
  The colors are shown in the legend of Fig.~\ref{fig:sigmatilde}.
}
\label{fig:ssum}
\end{center}
\end{figure}

\begin{figure}
\begin{center}
\includegraphics[width=0.7\linewidth]{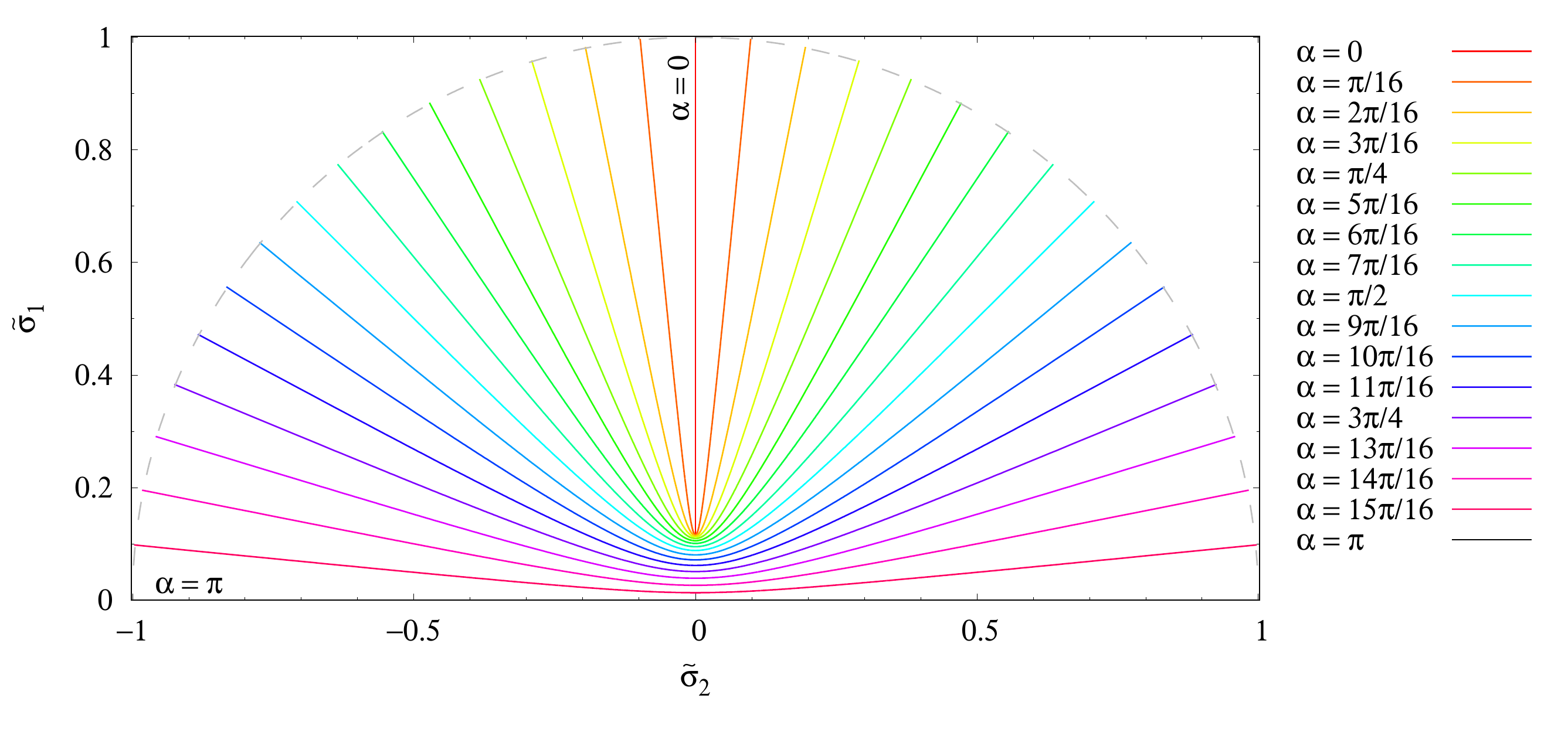}
\caption{\small  The functions  ${\tilde \sigma}_1(x)$ and
  ${\tilde\sigma_2}(x)$ are here shown in a parametric plot.  The
  $\mathbb{CP}^{N-1}$ coordinates are rotated so that the two
  boundaries look more symmetric, see Eq.~\eqref{genecond22} and
  Fig.~\ref{New4}.   ${\tilde\sigma}_{1,2}$  take values on the
  semicircle, i.e., points on $\mathbb{CP}^{N-1}$, at the
  boundaries.
  The gray dashed line drawing a semi-circle is the numerical cutoff,
  where the radius is given by $\sqrt{r_{n_{\rm max}}}$, see
  Eq.~\eqref{eq:rnmax}.
} 
\label{fig:sigmatilde}
\end{center}
\end{figure}

The numerical solutions to $\sigma_1(x)$ and $\sigma_2(x)$ for
$\alpha=\frac{n\pi}{16}$ with $n=0,1,2,\ldots,16$ are given in
Fig.~\ref{fig:sigmas}. 
In order to distinguish the solutions for different alpha, we
introduced a color scheme used throughout the paper, which is defined
as $2\alpha$ being mapped to the color circle\footnote{$2\alpha$ is
  mapped to the color angle called the hue. In
  our convention: $\alpha=0$ is red; $\alpha=\frac{\pi}{3}$ is green;
  $\alpha=\frac{2\pi}{3}$ is blue and the anti-colors are in between:
  $\alpha=\frac{\pi}{6}$ is yellow; $\alpha=\frac{\pi}{2}$ is cyan and
  finally, $\alpha=\frac{5\pi}{6}$ is magenta. }.
In order to distinguish $\alpha=0$ and $\alpha=\pi$, we changed the
color for the latter to black. 
For a legend with all the colors, see Fig.~\ref{fig:sigmatilde}.
These figures are consistent with the boundary conditions imposed,
i.e.~Eqs.~\eqref{genecond1} and \eqref{genecond2}.  An asymmetric
appearance  of $\sigma_1$ and $\sigma_2$ is due to the choice of the 
$\mathbb{CP}^{N-1}$ coordinates (the first line of
\eqref{generalbc}). 
The combination, $\sigma_1^2+\sigma_2^2$, which appears in the gap
equation \eqref{gapeq1} is shown in Fig.~\ref{fig:ssum}.
As $\sigma_1^2+\sigma_2^2$ is invariant under the rotations in the
$1-2$ plane in $\mathbb{CP}^{N-1}$, the picture is indeed
symmetric under the exchange of the two boundaries.    As can be seen
clearly, it shows a very little dependence on $\alpha$, except in the
middle region of the string.    In particular the value
$\sigma_1=\sigma_2=0$ is reached only for $\alpha=\pi$ and only at the
center of the string, $x=0$.
This corresponds to $\sigma_2\equiv 0$ on the entire string and
$\sigma_1$ crossing through zero at the midpoint.

\begin{figure}[!ht]
\begin{center}
\mbox{\subfloat{\includegraphics[width=0.49\linewidth]{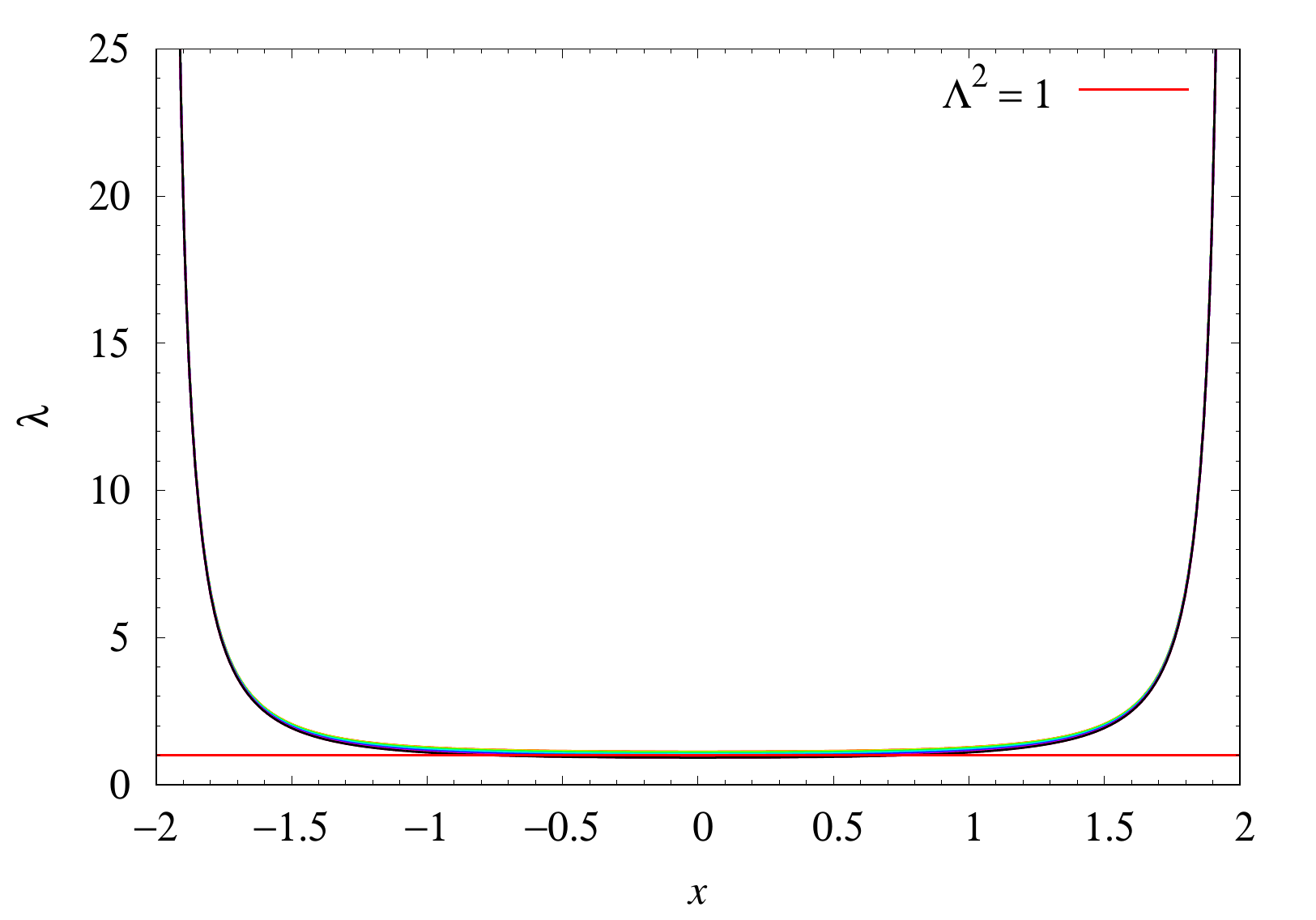}}
\subfloat{\includegraphics[width=0.49\linewidth]{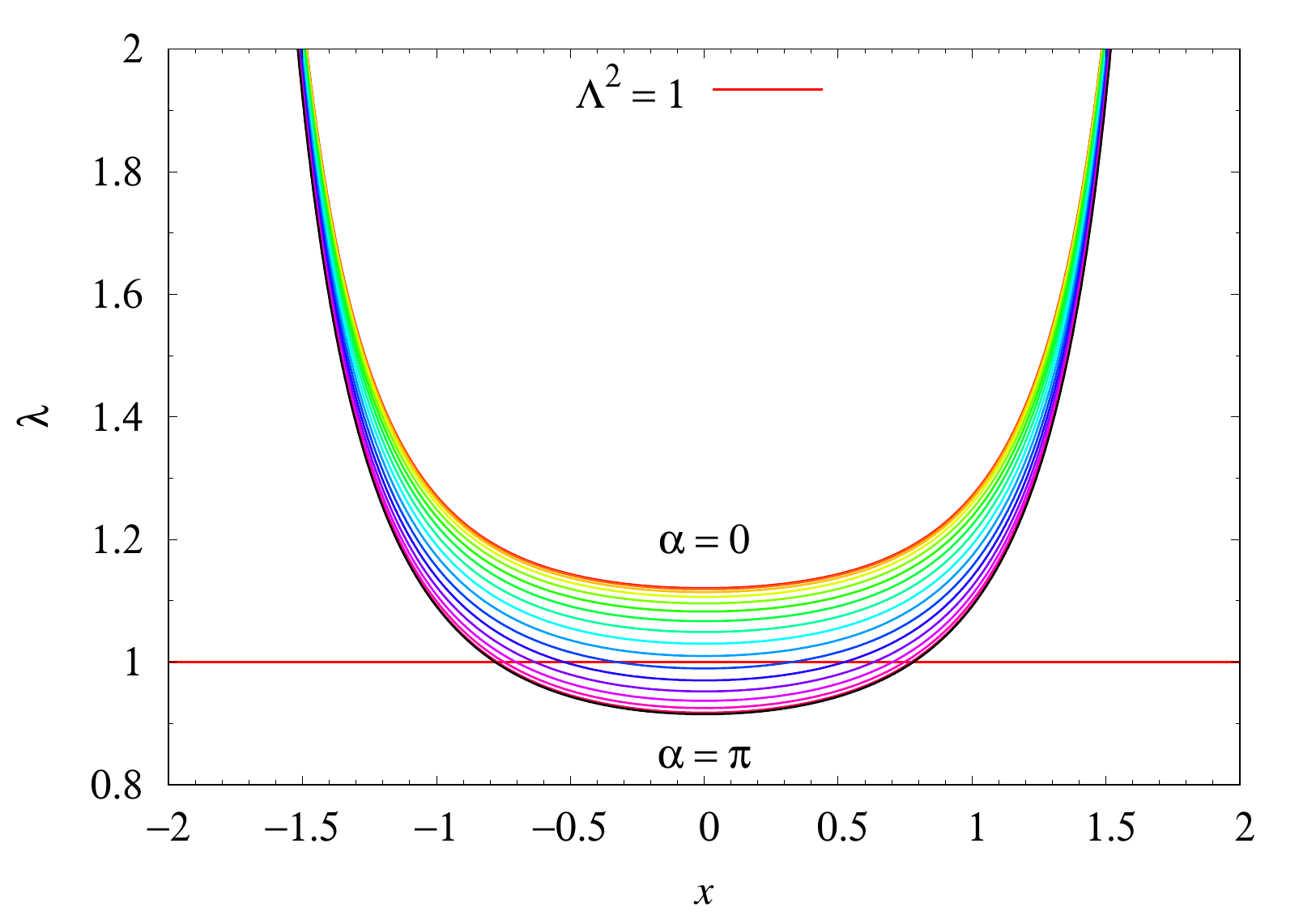}}}
\caption{\small The gap function  $\lambda(x)$ which solves the gap
  equation, Eq.~\eqref{gapeq1}, is plotted for various values of
  $\alpha$ for  $L=4$ and  $\Lambda=1$.
  $\alpha=0,\tfrac{\pi}{16},\tfrac{2 \pi}{16}, \ldots, \pi$ from the
  top curve to the bottom.
  On the right are the same curves zoomed in,  on the vertical.
  The colors are shown in the legend of Fig.~\ref{fig:sigmatilde}.
}
\label{fig:lambda}
\end{center}
\end{figure}

The gap function $\lambda(x)$ is plotted in Fig.~\ref{fig:lambda}.
Consistently with the previous Fig.~\ref{fig:ssum}, the gap function
depends on $\alpha$ significantly only in the central region of the
string.

In order to better see the symmetric nature of the two boundaries, one
can use the form of the boundary condition  (\ref{genecond11}),
(\ref{genecond22}) (see Fig.~\ref{New4}).  The parametric plot of the
solutions in terms of $\tilde {\sigma}_{1,2}$ is shown in
Fig.~\ref{fig:sigmatilde}.

\begin{figure}[!ht]
\begin{center}
\mbox{\subfloat{\includegraphics[width=0.4\linewidth]{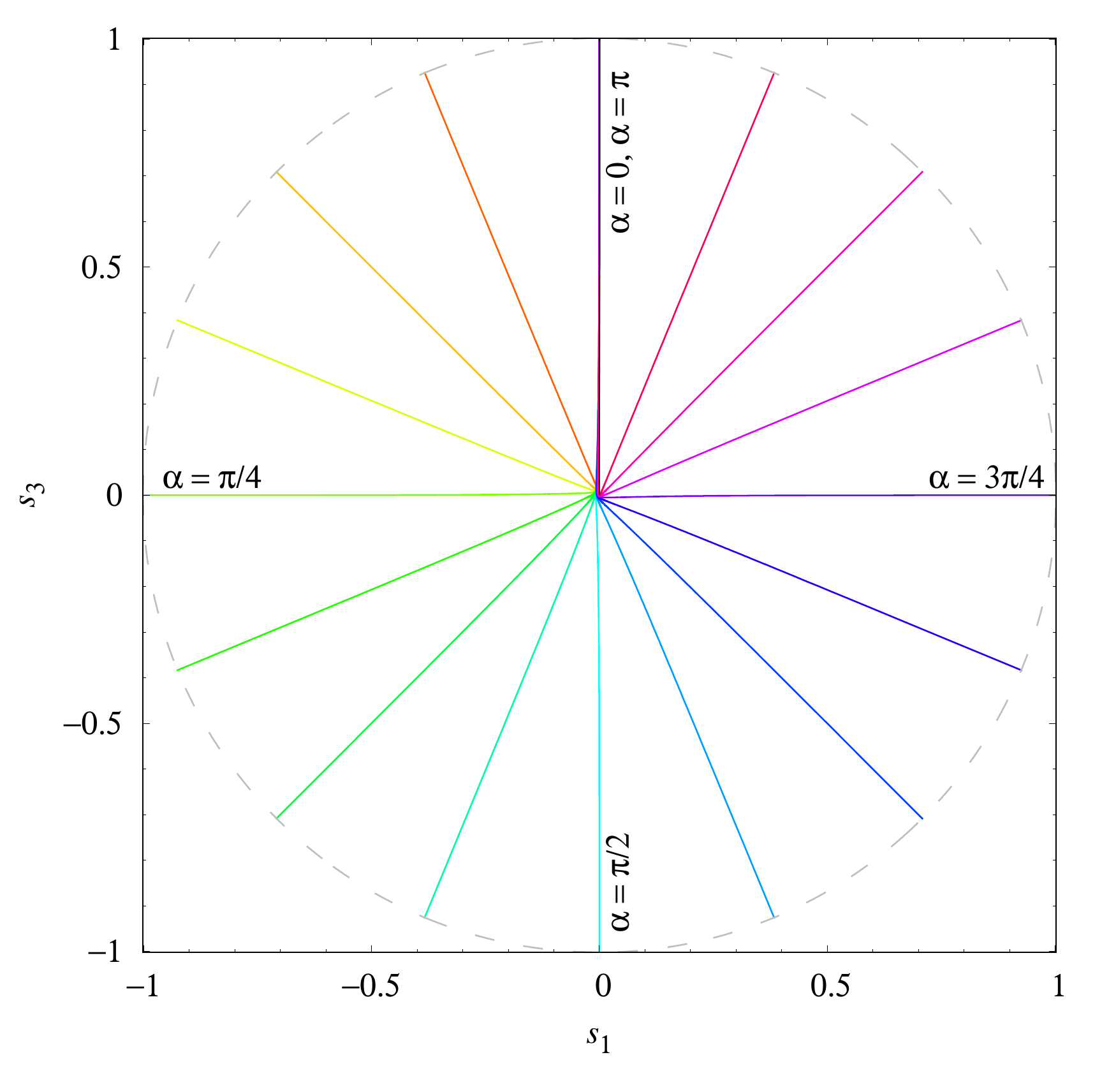}}
\subfloat{\includegraphics[width=0.4\linewidth]{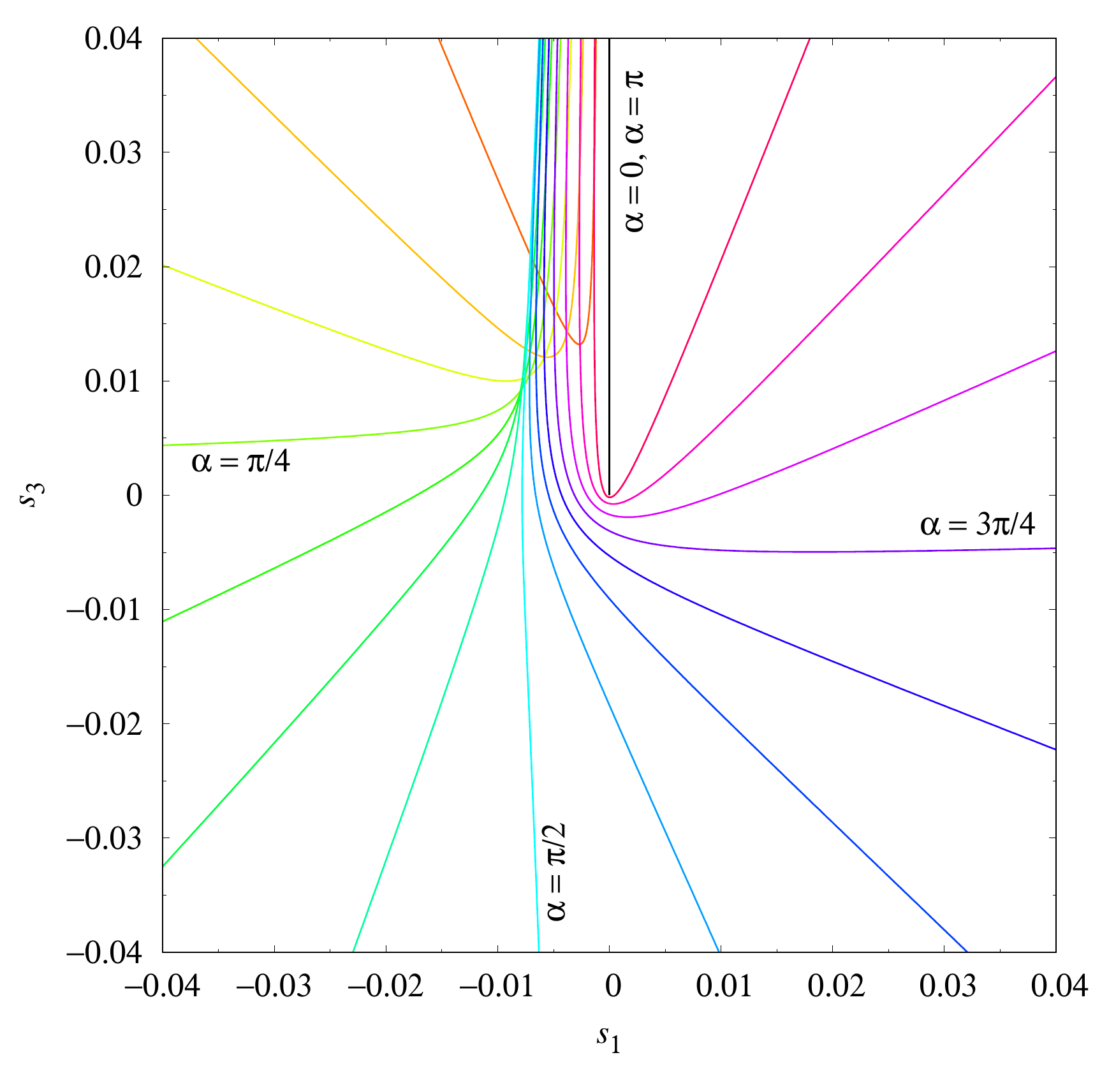}}}
\caption{\small The parametric plot of the solutions with the
  $\mathbb{CP}^{N-1}$ variables   $s_i$. On the right is the zoomed-in
  picture of the same curves.
  The gray dashed circle is the numerical cutoff, where the radius is
  given by $r_{n_{\rm max}}$, see Eq.~\eqref{eq:rnmax}.
  The colors are shown in the legend of Fig.~\ref{fig:sigmatilde}.
}
\label{fig:O3}
\end{center}
\end{figure}

These results  contain several interesting features which are not
always manifest. To reveal some of  them,  we introduce the
$\mathbb{CP}^{N-1}$ variables rather than the homogeneous coordinates
$n_i$. As only two components $n_{1,2}$ ($\sigma_{1,2}$) are involved
in the solution, it is sufficient to use the $\mathbb{CP}^{1}= S^2$
variables related to them by 
\be
s_i = \boldsymbol{\sigma}^{\dagger} \tau^i \boldsymbol{\sigma}\, , \qquad
\boldsymbol{\sigma} \equiv
\begin{pmatrix}
  \sigma_1\\
  \sigma_2
\end{pmatrix}\;,
\ee
where $\tau^i$ are the Pauli matrices. 
Also, as in the solutions considered $n_i$ are real, one may restrict
oneself to 
\be       s_3= \sigma_1^2 - \sigma_2^2 \;, \qquad   s_1= 2  \sigma_1 \sigma_2\;, \qquad s_2\equiv 0\;.
\ee
The solutions for $s_3$ and $s_1$ for several values of $\alpha$ are
given in a parametric plot in Fig.~\ref{fig:O3}. 

The analogous results for $L=1$ and $L=8$, corresponding to
Figs.~\ref{fig:sigmas}-\ref{fig:O3}, are collected in
Figs.~\ref{fig:sigmasBis}-\ref{fig:lambdaBis} and
Figs.~\ref{fig:sigmasBisBis}-\ref{fig:O3BisBis}, respectively,
of Appendix~\ref{sec:L1and8}.

\begin{figure}
\begin{center}
\includegraphics[width=3.5in]{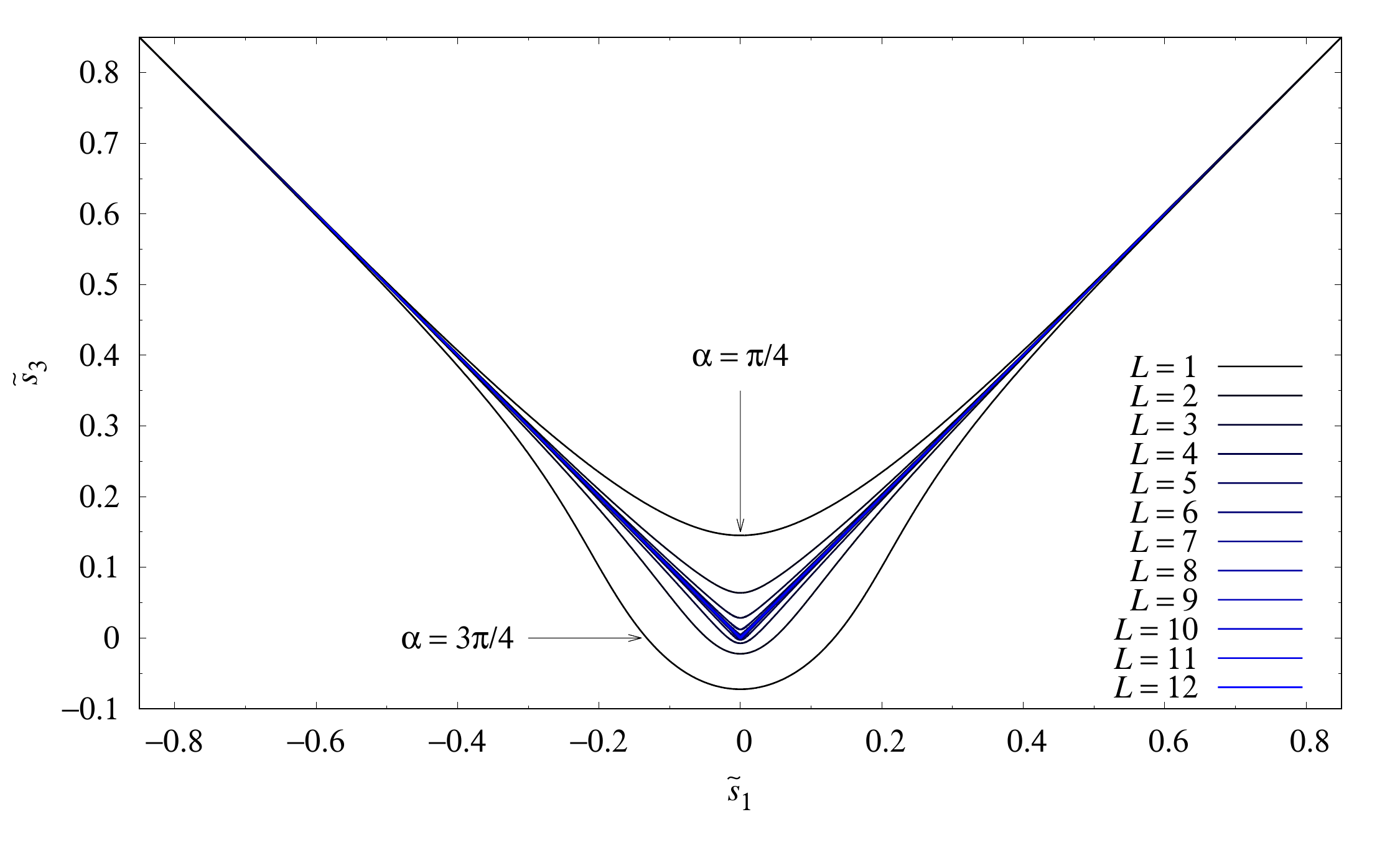}
\caption{A parametric plot of the ``symmetric'' $\mathbb{CP}^{N-1}$
  variables for $\alpha=\frac{\pi}{4}$ (above the two diagonal
  straight lines emanating from the origin) and
  $\alpha=\frac{3\pi}{4}$ (below the same two diagonal lines), for
  $\Lambda=1$ and $L=1,2,3,\ldots,12$.
  The color scheme is shown in the legend and is unrelated to that
  defined for the angles $\alpha$. 
}
\label{fig:sigmaLdep}
\end{center}
\end{figure}

Finally, we define some alternative $\mathbb{CP}^{N-1}$ variables
\be
\tilde{s}_i \equiv
  \tilde{\boldsymbol{\sigma}}^{\dagger} \tau^i
  \tilde{\boldsymbol{\sigma}}\,,\qquad
\tilde{\boldsymbol{\sigma}} \equiv
  \begin{pmatrix}
    \tilde{\sigma}_1\\
    \tilde{\sigma}_2
    \end{pmatrix}\, ,
\ee
by using the more the ``symmetric'' boundary conditions
\eqref{genecond11} and \eqref{genecond22}; thus by using the 
$\tilde{\sigma}_i$'s given in Eq.~\eqref{geneconsigmatilde}. 
Using these variables, the  $L$ dependence of the solutions are
illustrated in Fig.~\ref{fig:sigmaLdep}, for fixed values of
$\alpha$,  $\alpha= \tfrac{\pi}{4}$ and $\alpha= \tfrac{3\pi}{4}$,
as an example of two solutions with $\alpha$ and $\pi-\alpha$, see
Subsection \ref{sec:range}.

It is seen from the figure that only for small values of $L$, the
parametric solution bends inwards (for $\alpha\leq\frac{\pi}{2}$) or
outwards (for $\alpha\geq\frac{\pi}{2}$).
As $L$ tends to infinity (for fixed and finite $\Lambda$), the
parametric solution tends to exactly the diagonals. That is, in these
coordinates, the solution will come in with an angle
$\frac{\pi}{2}-\alpha$ with respect to the $y$ axis and go out with an
angle $\alpha-\frac{\pi}{2}$. 

Indeed, as the two points $\alpha$ and $\pi-\alpha$ are the same point
in $\mathbb{CP}^{N-1}$, only one of the two solutions can be stable
and the other will be metastable.
The figure alludes to the claim that the solution with
$\alpha\leq\frac{\pi}{2}$ is the stable one; in this case, the
solution with $\alpha=\frac{\pi}{4}$, whereas the other is a
metastable solution.

In order to make evidence for our claim, we will study the $\alpha$
dependence of the total string energy in the next section.

\section{Dependence of the energy on the relative orientation $\alpha$}\label{sec:alpha}

Let us now analyze how the total energy:
\be   E  =   \int  dx  \left[   \frac{N}{2}  \sum_n \Big\{   \omega_n  f_n^2 + \frac{1}{\omega_n} ( f_n^{\prime \, 2} + \lambda f_n^2 ) \Big\} \,e^{-\epsilon \omega_n}       + \sigma_1^{\prime \, 2} + \sigma_2^{\prime \, 2}   +\lambda (\sigma_1^2 +\sigma_2^2 - r^0_{\epsilon}) \;
  \right] \;,
\ee
depends on the relative orientation $\alpha$.  
It is given by
\bea   \frac{\partial E}{\partial \alpha}   &=&  \int dx\, \left[  \frac{\partial \sigma_a(x)}{\partial \alpha}    \frac{ \delta E}{\delta \sigma_a(x) } +  
 \frac{\partial \lambda(x)}{\partial \alpha}    \frac{ \delta E}{\delta \lambda(x) }  \right]   \nonumber \\
 &=&  \left[  \frac{\partial \bar{\sigma}_a}{\partial \alpha} (\sigma_a^{\prime}) + \frac{\partial {\sigma}_a}{\partial \alpha} (\bar \sigma_a^{\prime}) \right]_{-L/2}^{L/2}
= 2  \left[   \frac{\partial {\sigma}_a}{\partial \alpha} (\sigma_a^{\prime}) \right]_{-L/2}^{L/2}  \; ,\label{kept}
     \eea
where      $\sigma_a$ is taken to be real and repeated indices are
summed over: $a=1,2$.
Note that  as $\lambda(x)$ and $\sigma_a$ satisfy the equation of motion, 
\be
\frac{\delta E}{\delta \lambda(x)}=0\;; \qquad
\frac{\delta E}{\delta \sigma_a(x)}=0\;,\quad a=1,2\;,
\ee
the  contribution from the interior of the string,
$(-\tfrac{L}{2}, \tfrac{L}{2})$,  vanishes,    and  only the surface
term remains.
We will now use the general solution, the first line of (\ref{generalbc}), to write
\bea 
\frac{\partial E}{\partial \alpha} &=&  2\left[ \frac{\partial \sigma_1}{\partial \alpha}  \sigma_1^{\prime} +    \frac{\partial \sigma_2}{\partial \alpha}  \sigma_2^{\prime} \right]_{-L/2}^{L/2}  \label{secondline}\\
   &=&     \left[
- 2  \sin \alpha \,    \sigma_L  \sigma_R'   \,\right]_{-L/2}^{L/2}    
+  2 \left[   \frac{\partial \sigma_R}{\partial \alpha} \sigma_R'  +    \frac{\partial \sigma_L}{\partial \alpha} \sigma_L'  +
\cos \alpha \left(    \frac{\partial \sigma_R}{\partial \alpha} \sigma_L'  + \frac{\partial \sigma_L}{\partial \alpha} \sigma_R' 
\right)
\right]_{-L/2}^{L/2}\;.  \nonumber   
 \eea
By using parity one can write this as  
\be
\frac{\partial E}{\partial \alpha} = 2\sin \alpha \, \left[    \sigma_L \sigma_R'  +   \sigma_R  \sigma_L'  \right]_{-\tfrac{L}{2}}   
+  4 \left[   \frac{\partial \sigma_R}{\partial \alpha} \sigma_R'  +    \frac{\partial \sigma_L}{\partial \alpha} \sigma_L'  +
\cos \alpha \left(    \frac{\partial \sigma_R}{\partial \alpha} \sigma_L'  + \frac{\partial \sigma_L}{\partial \alpha} \sigma_R' 
\right)
\right]_{-\tfrac{L}{2}}  \,,    \label{negligible}
\ee
where now all functions are  evaluated at the left boundary,
$x=-\tfrac{L}{2}$. \footnote{Naturally, this can be expressed in terms
  of the values of the functions at the right boundary instead, 
with an over all minus sign in front. The result is the same.}  

Let us first evaluate the second term of (\ref{negligible}) containing  $\frac{\partial \sigma_{L, R}}{\partial \alpha}$.   
By using the behavior of   $\sigma_{L,R}$  near the left  boundary, $x \sim -\tfrac{L}{2}$,  
\be   
   \sigma_L  \sim   \sqrt{ \tfrac{N}{2\pi}   \log \tfrac{1}{x + \frac{L}{2}}}\;, \qquad  \sigma_R   \sim W   \frac { x + \tfrac{L}{2}} {\sqrt{    \tfrac{N}{2\pi}  \log \tfrac{1}{x + \frac{L}{2} }}}  \;, \label{leading}\ee
one gets the estimates
\be     \lim_{x\to -\frac{L}{2} }  \frac{\partial \sigma_R}{\partial \alpha} \sigma_R'  =  \lim_{x\to -\frac{L}{2} }   \mathcal{O}
\left( \frac { x + \tfrac{L}{2}} {   \tfrac{N}{2\pi}  \log \tfrac{1}{x + \frac{L}{2} }} \right)=0\;;
\ee
\be   \lim_{x\to -\frac{L}{2}}   \frac{\partial \sigma_R}{\partial \alpha} \sigma_L'  =  \lim_{x\to -\frac{L}{2} }   \mathcal{O}
\left( \frac { 1} { \tfrac{N}{2\pi}  \log \tfrac{1}{x + \frac{L}{2} }  }  \right)=0\;.
\ee
The evaluation of terms involving $\frac{\partial \sigma_{L}}{\partial \alpha}$ requires a more careful consideration:  although the leading behavior of $\sigma_L$  in  (\ref{leading}) is independent of $\alpha$,  the $\alpha$ dependent effects coming from subleading terms may not be negligible. 
A study in Appendix~\ref{sec:Care}, however, shows that  
\begin{eqnarray}
\lim_{x \to-  \frac{L}2} 
\frac{\partial \sigma_{\rm L}(x)}{\partial \alpha} \sigma_{\rm L}'(x) = \lim_{x \to - \frac{L}2}    \frac{\partial \sigma_{\rm L}(x)}{\partial \alpha} \sigma_{\rm R}'(x)=
\mathcal{O} \left( \frac1{  \tfrac{N}{2\pi}   \log (\frac{L}2 + x)}\right)=0\;.  \label{evaluation}
\end{eqnarray}


As for the first term of Eq.~\eqref{negligible}, it can be evaluated
straightforwardly: 
\begin{eqnarray}
\frac{\partial E(\alpha)}{\partial \alpha}&=& 2 \sin \alpha  \left[ \sigma_{\rm R}\left( -\tfrac{L}2\right)\sigma_{\rm L}'\left(- \tfrac{L}2\right)  
+\sigma_{\rm L}\left( -\tfrac{L}2\right)\sigma_{\rm R}'\left(- \tfrac{L}2\right)  \right]\nn
&=& 2 \sin \alpha  \left[W +2\sigma_{\rm R}\left( -\tfrac{L}2\right)\sigma_{\rm L}'\left( -\tfrac{L}2\right)  \right],
\end{eqnarray}
where again the definition of the Wronskian \eqref{Wrnsk} has been used.
As
\begin{eqnarray}
  \lim_{x\to -\frac{L}2}\sigma_{\rm R}(x)\sigma_{\rm L}'(x)= \lim_{x\to- \frac{L}2} {\cal O}\left(\frac 1{-\log (\frac{L}2+x)}\right)=0\;,
\end{eqnarray}
one gets  the net result
\be
\frac{\partial E}{\partial\alpha} =  +  2  \sin \alpha \, W >0\;, \qquad  0< \alpha < \pi\;.   \label{deEdealpha}
\ee
The energy increases monotonically with $\alpha$. 

\begin{figure}
\begin{center}
\includegraphics[width=3.5 in]{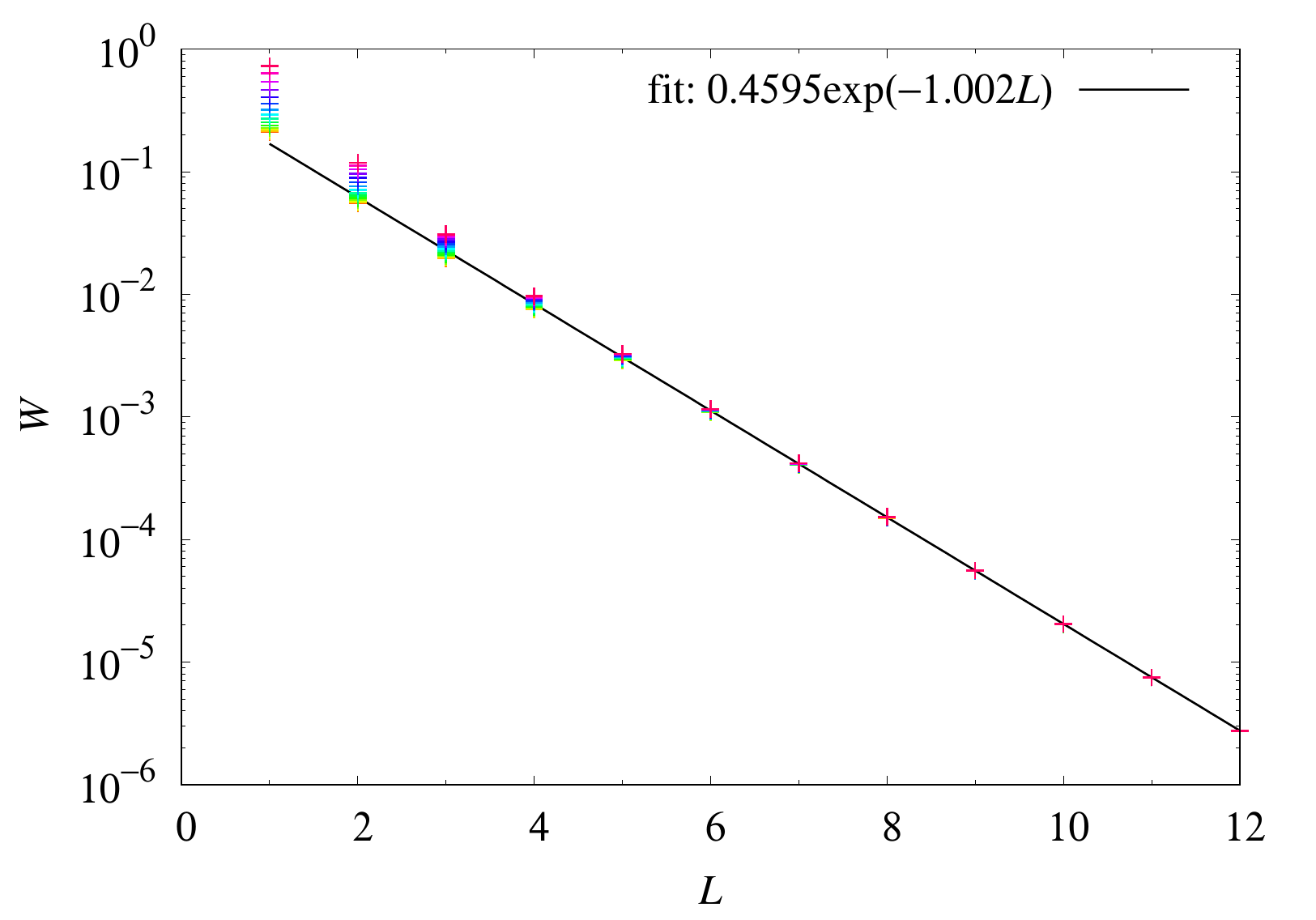}
\caption{\small The Wronskian  (\ref{Wrnsk}) is numerically evaluated
  for various $\alpha$, as a function of $L$ for $\Lambda=1$.
  The colors are shown in the legend of Fig.~\ref{fig:sigmatilde}.
} 
\label{fig:wronskian}
\end{center}
\end{figure}

Even though  at large $L$  the Wronskian  $W$   is believed to be exponentially small
 \footnote{This is so, as  both $\sigma_R$ and $\sigma_L$  behave as $\sim e^{- L \Lambda / 2}$, at distance $\tfrac{L}{2}$ from the boundaries.  See  Sections 5 and 6 of Ref.~\cite{BBGKO}.  },
 $\mathcal{O}(e^{-L \Lambda})$,
  so the $\alpha$ dependence of $E$ is very small,   this is not so at smaller $L\sim \mathcal{O}(1/\Lambda).$  One expects a significant dependence of the energy on $\alpha$. 
See Fig.~\ref{fig:wronskian} for the numerically found  $W$ for various values of $\alpha$, as a function of $L$.   One sees that the exponential $L$
dependence at large $L$ is universal, i.e., independent of $\alpha$,
which is quite understandable,  as the effect of mis-alignment at the
$\mathbb{CP}^{N-1}$ variables at the far boundaries should be
unimportant at large $L$.

Eq.~(\ref{deEdealpha}) is one of the main results of the present work:
it   proves that, in the $U(1)$ gauge in which the $\sigma_{1,2}(x)$
fields are real throughout $[-\tfrac{L}{2}, \tfrac{L}{2}]$ and
$A_x \equiv 0$, the solutions in the range
$0 \le   \alpha   \le  \frac{  \pi }{2} $    (Eq.~\eqref{range}) are
the stable ones, as claimed.

\subsection{Numerical checks of  $\frac{\de E}{\de \alpha}$}

The total energy of the system can be expressed in various ways:
\begin{align}
  E &=   \int  dx  \left[   \frac{N}{2}  \sum_n \Big\{   \omega_n
  f_n^2 + \frac{1}{\omega_n} ( f_n^{\prime \, 2} + \lambda f_n^2 ) \Big\}
  e^{-\epsilon \omega_n}       + \sigma_1^{\prime \, 2} +
  \sigma_2^{\prime \, 2}   +\lambda \left(\sigma_1^2 +\sigma_2^2 -
  r^0_{\epsilon}\right) 
\right]  \label{first}\\
  &= N \sum_n \omega_n e^{-\epsilon\omega_n}
  +\int  dx  \left[\sigma_1^{\prime \, 2} + \sigma_2^{\prime \, 2}   +\lambda (\sigma_1^2 +\sigma_2^2 - r^0_{\epsilon}) 
  \right],\label{Keisuke}
\end{align}
where we have integrated $f_n^{'2}$ by parts, used the fact that the
boundary term vanishes \cite{BBGKO}, the equation of motion for the
modes $f_n$ and the completeness relation \eqref{operator}.
The latter expression was the starting point in
Section~\ref{sec:solutions} and defining
\be
G \equiv  \int dx  \left( \sigma_1^{\prime \, 2}  +
\sigma_2^{\prime \, 2} \right)\;,    
\ee
it can be written neatly as
\be
E = N \sum_n \omega_n e^{-\epsilon\omega_n}
+ G 
+ \int dx \; \lambda \left(\sigma_1^2 +\sigma_2^2 -
r^0_{\epsilon}\right).
\ee
Starting again from the first expression \eqref{first}; instead of
integrating by parts, we can collect the terms that are multiplied by
the gap function $\lambda$:
\begin{align}
 E &= \int  dx  \bigg[  \left( \frac{N}{2}  \sum_n \Big\{   \omega_n
   f_n^2 + \frac{1}{\omega_n} f_n^{\prime \, 2} \Big\} e^{-\epsilon
     \omega_n}   + \sigma_1^{\prime \, 2} + \sigma_2^{\prime \, 2}
   \right)  \nonumber\\
   &\phantom{= \int dx \bigg[}
   +  \lambda  \left(   \frac{N}{2}    \sum_n   \frac{f_n^2 }{\omega_n}  e^{-\epsilon \omega_n}     +      \sigma_1^2 +\sigma_2^2 - r^0_{\epsilon}  \right)\bigg] 
 \nonumber   \\
 &=  \int dx  \,  \left[   {\cal E}_0  +     \frac{N}{2\pi}  \lambda
   \right] 
 = {\cal E}_0  \, L +      \frac{N}{2\pi}   \int dx  \lambda       \;. \label{last}
\end{align}
where 
\be   {\cal E}_0 \equiv    \frac{N}{2}  \sum_n \Big\{   \omega_n  f_n^2 + \frac{1}{\omega_n} f_n^{\prime \, 2} \Big\} e^{-\epsilon \omega_n}   + \sigma_1^{\prime \, 2} + \sigma_2^{\prime \, 2}  \,,
\label{eq:calE0def}
\ee
can be shown to be a constant  (independent of $x$),  by using the gap
equation \cite{BBGKO}.
Also the  equation of motion of $\lambda$
\be   \frac{N}{2}  \sum_n   \frac{f_n^2 }{\omega_n}  e^{-\epsilon \omega_n}     +      \sigma_1^2 +\sigma_2^2 =    r_{\epsilon},  \qquad    r_{\epsilon} =   r^0_{\epsilon}  +  \frac{N}{2\pi}\,,
\ee
has been used in the last step.  

 As   the equations of motion for $\sigma_a$ and  $\lambda$,  Eq.~(\ref{gapeqbbGen1}) and  Eq.~(\ref{gapeqbbGen2}) have been  used in moving among the lines,  these expressions are equivalent on-shell, i.e.,  when evaluated at the minimum of the action.     For the purpose of numerically verifying  the $\alpha$ dependence of the energy $E$ found analytically above, however,  use of different expressions will provide us with 
nontrivial, independent checks.  

A particularly subtle issue in our discussion concerns the divergences.  The constant part of the energy density, ${\cal E}_0$,  is quadratically divergent, even after the logarithmic divergence in the sum over modes  is eliminated by the standard coupling constant renormalization;   the subtraction prescription  \cite{BBGKO}     
\be        {\cal E}_{\rm uv}      = -  \frac{N}{\pi \epsilon^2} \; , \label{subtract}
\ee
is implicit   in Eqs.~(\ref{first})-(\ref{last}).   Moreover,  due to the behavior of the gap function 
\be
\lambda(x) \sim \frac{1}{2\left|x\pm \tfrac{L}{2}\right|^2\log \tfrac{1}{\left|x\pm L/2\right| \Lambda}} \;, \qquad
L \gg  1/\Lambda  \gg  |x\pm \tfrac{L}{2}|  \;,   
\ee
near the boundaries,  integration of the gap function in Eq.~(\ref{last})
diverges linearly, which should be canceled by the bare
``monopole mass'' terms \cite{BBGKO}.  
Both the quadratic divergence of the energy density and the linear
divergence of the $\lambda$ integration are {\it  local} effects,  the
former around a generic point in the string, $x$, and the latter at
the boundaries;  therefore, they should be independent of the relative
$\mathbb{CP}^{N-1}$ orientation $\alpha$ at the two boundaries.
The derivation of Eq.~\eqref{deEdealpha} relies on this tacit
assumption: it is a highly nontrivial check whether this result is
reproduced by the evaluation of some of
Eqs.~\eqref{first}-\eqref{last}, by inserting the numerical solutions
of the gap equation discussed in Section~\ref{sec:solutions} and
illustrated in Figs.~\ref{fig:sigmas}-\ref{fig:O3}.

For the numerical check,  a possibility is to use the formula 
\bea  \frac{\de E}{\de \alpha}   &=&     \frac{\de G }{\de \alpha} +   \int dx    \,    2  \lambda   \left(   \sigma_1  \frac{\de \sigma_1}{\de \alpha } +   \sigma_2  \frac{\de \sigma_2}{\de \alpha } \right) \nonumber \\
&=&   2  \int  dx  \, \left[  \left(\sigma_1^{\prime} \frac{\de \sigma_1^{\prime}}{\de  \alpha} + \sigma_2^{\prime} \frac{\de \sigma_2^{\prime}}{\de  \alpha} \right) +    \lambda   \left(   \sigma_1  \frac{\de \sigma_1}{\de \alpha } +   \sigma_2  \frac{\de \sigma_2}{\de \alpha } \right) \right], \label{check}
\eea
which follows from differentiating Eq.~\eqref{Keisuke} with respect to
$\alpha$.  The dependence on $\alpha$ through the function
$\lambda(x)$, proportional to the gap equation \eqref{gapeqbbGen1},
has been dropped (i.e., made use of), hence Eq.~\eqref{check}
expresses
$\de E/\de \alpha $  through the functional dependence on $\sigma_a$
only.
If the equations of motion for $\sigma_a$ (Eq.~\eqref{gapeqbbGen2})
were also used, the only thing that remains would be the surface terms
-- which have been evaluated analytically in Section~\ref{sec:alpha}.

Another possibility is to use Eq.~\eqref{last} instead: 
\be  \frac{\de E}{\de \alpha} = L\,   \frac{\de {\cal E}_0 }{ \de \alpha }  +   \frac{N}{2\pi}      \int dx    \,  \frac{\de \lambda}{\de \alpha}  \;,      \label{checkBis}
\ee
where the constancy of $ {\cal E}_0$ can be exploited to evaluate the
first term  $ \tfrac{\de {\cal E}_0 }{ \de \alpha }$, e.g., at the
midpoint of the string, $x=0$, where the numerical precision is best.

\begin{figure}[!ht]
\begin{center}
\mbox{\subfloat{\includegraphics[width=0.49\linewidth]{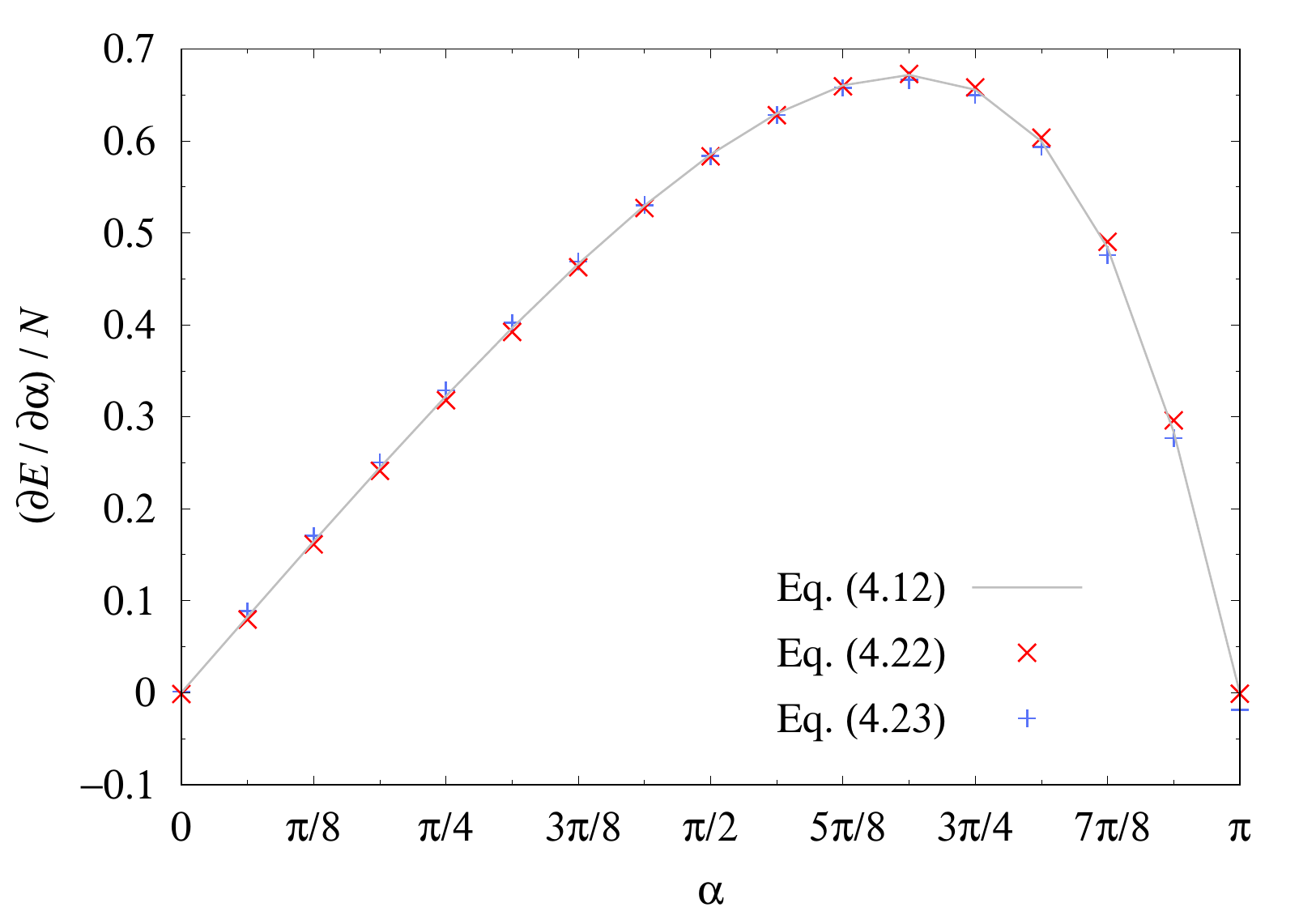}}
\subfloat{\includegraphics[width=0.49\linewidth]{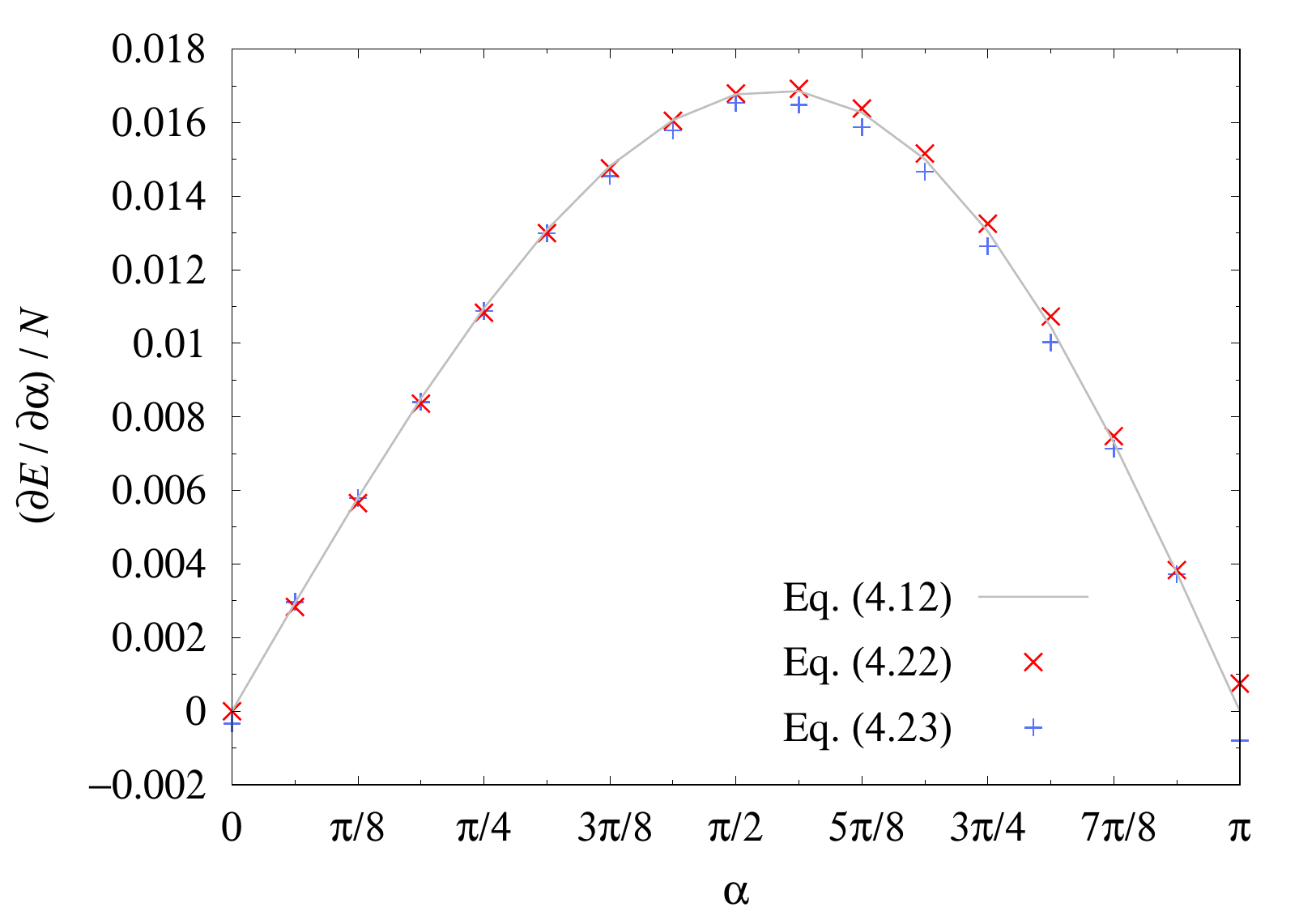}}}
\caption{\small The $\alpha$ dependence of the total energy $E$ is numerically  
evaluated by using Eq.~(\ref{check}) (red crosses)   and  Eq.~(\ref{checkBis}) (blue plus symbols), and compared to the analytic expression  Eq.~(\ref{deEdealpha})  (the solid line), as a function of $\alpha$.  $L=1$, $\Lambda=1$ in the left figure;    
$L=4$, $\Lambda=1$  in the right figure,  and $\alpha \in [0, \pi]$ in both.   
}
\label{fig:dEdA}
\end{center}
\end{figure}

The result of such a comparison is shown in Fig.~\ref{fig:dEdA},  for
$L=1$ and $L=4$  (and $\Lambda=1$).  Shown in the Figure are the
numerical evaluations of Eqs.~\eqref{check} and \eqref{checkBis} and
the analytical expression
$  \partial E / \partial\alpha=+2\sin\alpha \, W$,  where the
Wronskian is calculated numerically (Fig.~\ref{fig:wronskian}).
Apart from small numerical errors, the overall agreement between the
direct numerical evaluation and  the analytic formula is quite
satisfactory.
They clearly confirm our conclusion that the solutions in the range
$\alpha \in [0, \tfrac{\pi}{2}]$ are the stable ones.

\section{Discussion}\label{sec:discussion}

In this work we have further examined the quantum vacuum configuration of the  bosonic $\mathbb{CP}^{N-1}$ model \cite{DAdda:1978vbw,Witten:1978bc}, defined on finite space interval $L$,  i.e., on a finite-width worldstrip, in the large $N$ approximation.   
Building on the results of our preceding work \cite{BKO,BBGKO}, the systems with two  generic $\mathbb{CP}^{N-1}$ 
orientations at the two boundaries are studied here, for various values of the string length $L$ and for different relative orientation angle $\alpha$.  
The total energy of the system $E$ at fixed $L$  is found to increase with $\alpha$ monotonically.  Taking into account of the 
defining properties of the $\mathbb{CP}^{N-1}$ sigma model, this means that the relative angle between the classical field orientations
at the boundaries can be limited to $[0, \tfrac{\pi}{2}]$, without
loss of generality.   The system has the lowest energy when the $\mathbb{CP}^{N-1}$ orientation is the same at the two boundaries. 

The classical field component, in fact, traces a path from a point to
another in  $\mathbb{CP}^{N-1} \times \mathbb{R}_{>0}$, in going from
the left boundary to the right boundary.  It is on a point in
$\mathbb{CP}^{N-1}$ at the left boundary; it goes into  its
``interior'', before emerging at another point of the
$\mathbb{CP}^{N-1}$ surface at the other boundary (see
Fig.~\ref{New4}).
Due to the fact that the space
$\mathbb{CP}^{N-1} \times\mathbb{R}_{>0}$ is simply connected, the
solution with $\alpha$ and the one with $\pi - \alpha$ 
correspond to two paths which are homotopic to each other, as
discussed in Subsection \ref{sec:range}.
Thus only one of them can be stable.  The fact that
the energy $E$ monotonically increases with $\alpha$ in
$\alpha \in [0, \pi]$ shows that the stable solutions correspond to
those with $\alpha\in[0, \tfrac{\pi}{2}]$.

In this discussion the specific property of the $\mathbb{CP}^{N-1}$
model, not shared by other sigma models such as the $O(N)$ model,
turned out to be crucial. 

\begin{figure}[!ht]
\begin{center}
\mbox{\subfloat{\includegraphics[width=0.49\linewidth]{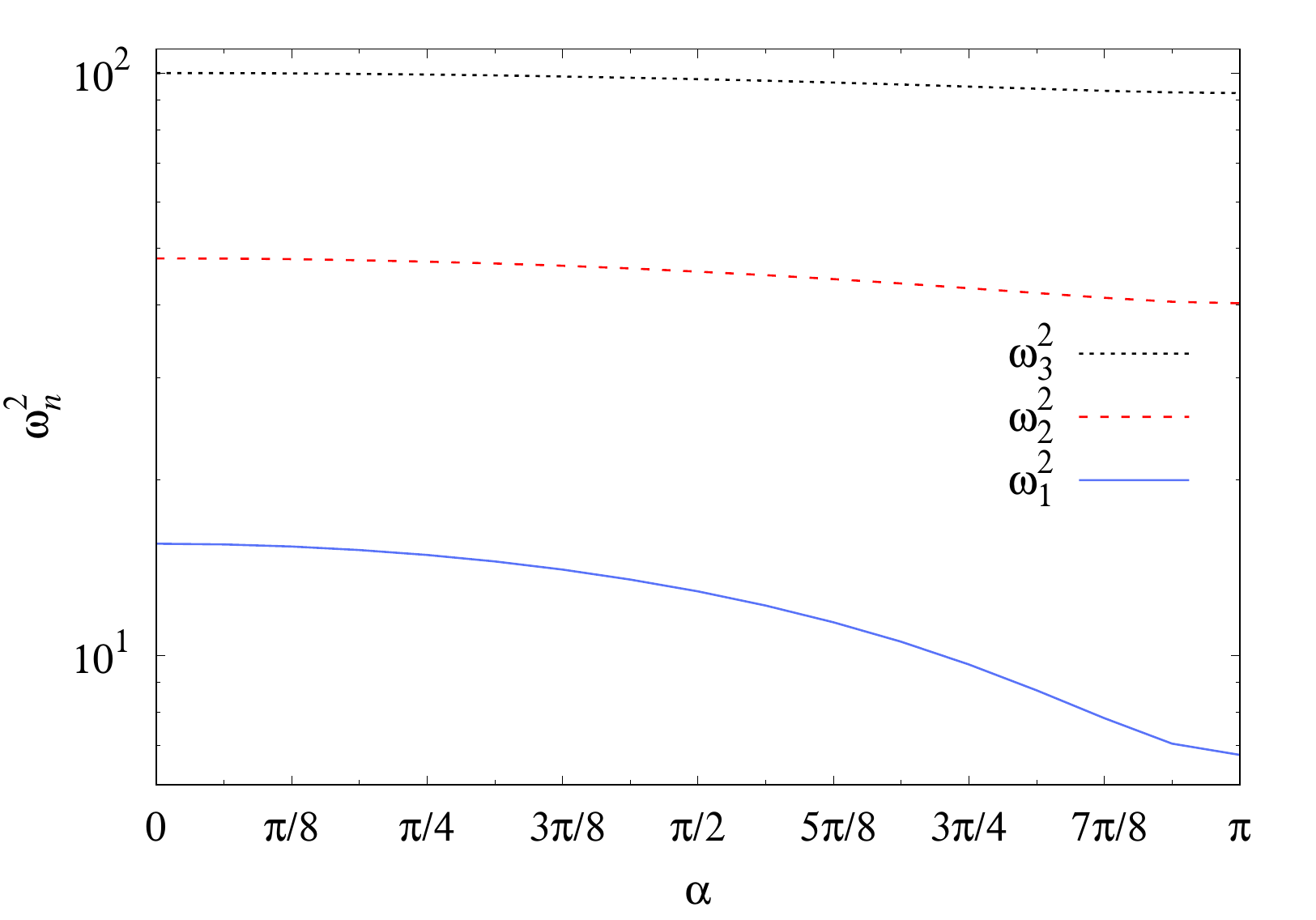}}
\subfloat{\includegraphics[width=0.49\linewidth]{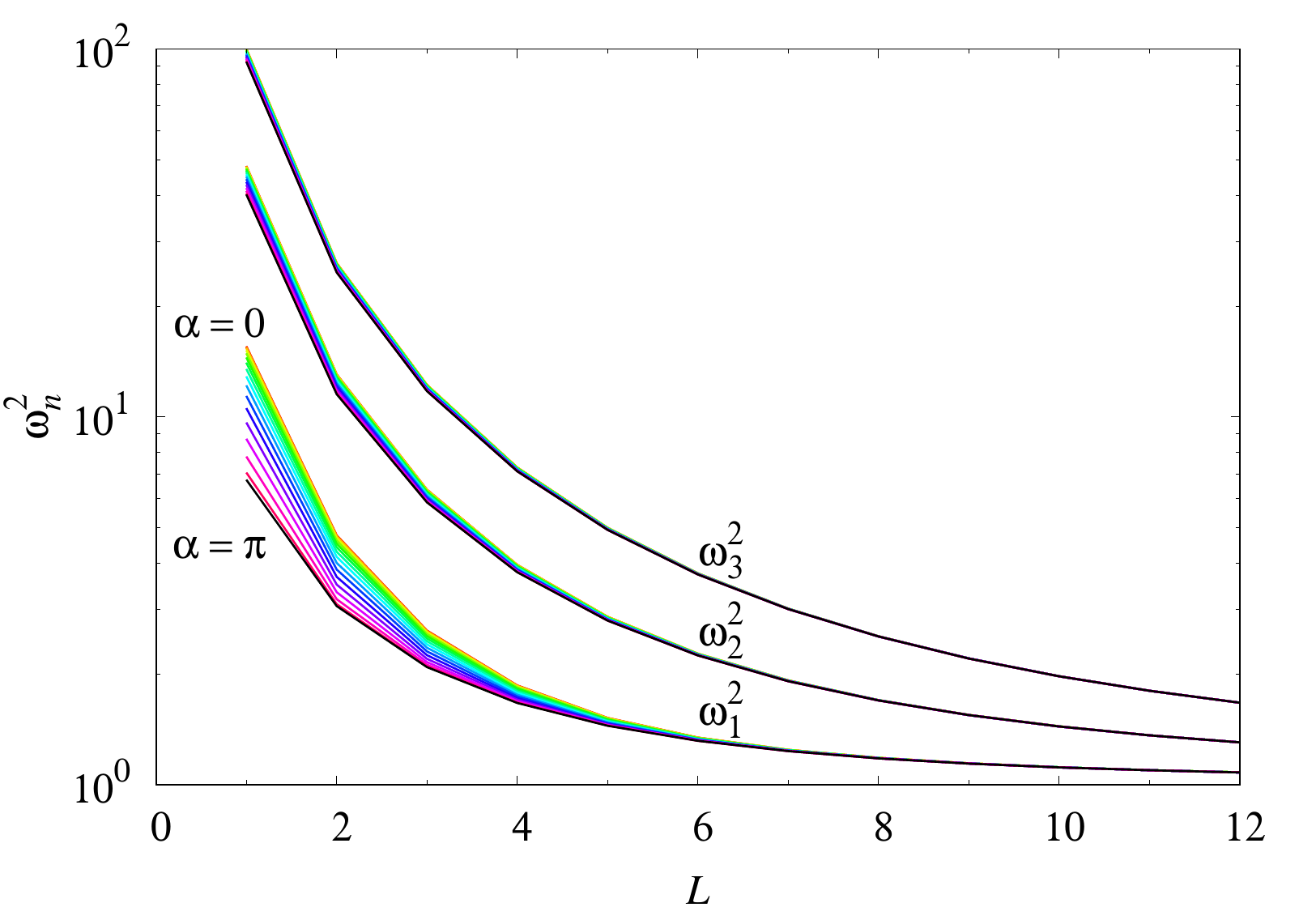}}}
\caption{\small The first three energy levels $\omega_n^2$ ($n=1,2,3$)
  of  (\ref{operator}),  calculated with potential $\lambda(x)$, for
  $L=1$ and as functions of $\alpha$ (on the left), and,    for
  various $\alpha$,  as functions of $L$ (on the right).
  The colors used in the right panel are shown in the legend of
  Fig.~\ref{fig:sigmatilde}.
}
\label{fig:Spectrum}
\end{center}
\end{figure}

The instability of the solutions with
$\alpha \in [\tfrac{\pi}{2},\pi]$ cannot be seen perturbatively.  We
have indeed verified that our ``solutions'' with
$\alpha > \tfrac{\pi}{2}$  do not suffer from any zero or negative
modes $\{f_n, \omega_n\}$  even if the potential $\lambda(x)$ becomes 
negative in the central region of the string  (the left of
Fig.~\ref{fig:lambdaBis}), see Fig.~\ref{fig:Spectrum}.
The instability is a nonperturbative phenomenon. 
It would be an interesting problem to understand better the nature of 
such instabilities.

\section*{Note added}

After this work is finished, we were informed by Muneto Nitta of their new paper \cite{Nitta:2018lnn}
which deals with a similar system.  As far as we can see our approach and the results obtained are different from theirs.

\section*{Acknowledgment}

The work of S.~B.~is funded by the grant ``Rientro dei Cervelli Rita
Levi Montalcini'' of the Italian government. 
The work of S.~B.~G.~is supported by the National Natural Science
Foundation of China (Grant No.~11675223).
K.~O.~is supported by the Ministry of Education, Culture, Sports,
Science (MEXT)-Supported Program for the Strategic Research Foundation
at Private Universities “Topological Science” (Grant No.~S1511006) and
by the Japan Society for the Promotion of Science (JSPS) Grant-in-Aid
for Scientific Research (KAKENHI Grant No.~16H03984).
This work is supported by the INFN special project grant ``GAST'' (Gauge
and Strong Theories).
We thank M. Nitta for informing us about their new paper.

\appendix

\section{Evaluation of Eq.~(\ref{evaluation})}\label{sec:Care}

The $\alpha$ dependence in $\sigma_{\rm R,L}(x)$ comes only  through
its functional dependence on  $\lambda(x)$: 
\begin{eqnarray}
\frac{ \partial \sigma_{\rm R,L}(x)}{\partial \alpha} = \int  dy \frac{\partial \lambda(y)}{\partial \alpha} \frac{\delta\sigma_{\rm R,L}(x)}{\delta \lambda(y)},   
\end{eqnarray}
which can be cast in a local differential equation form:
\begin{eqnarray}
\left( \partial_x^2-\lambda(x)\right) \frac{ \partial \sigma_{\rm R,L}(x)}{\partial \alpha}
=\frac{\partial \lambda(x)}{\partial \alpha} \sigma_{\rm R,L} (x). \label{used1} 
\end{eqnarray}
Let us consider the following quantity
\begin{eqnarray}
\Delta_\alpha \sigma_{\rm R,L}(x)\equiv \frac{\partial \sigma_{\rm R,L}}{\partial \alpha}(x) - \sum_{n} b_n^{\rm R,L}  f_n(x),   \label{used2} 
\end{eqnarray}
with a set of coefficients $\{ b_n^{\rm R, L}\} $ given by
\begin{eqnarray}
b_n^{\rm R,L} = -\frac1{\omega_n^2} \int_{-\frac{L}2}^{\frac{L}2} dy  \frac{\partial \lambda (y)}{\partial \alpha} \sigma_{\rm R,L}(y) f_n(y)\;.   \label{used3} 
\label{eq:bn}
\end{eqnarray}
There are relations between coefficients as  $b_n^{\rm R}=(-1)^n b_n^{\rm L}$ due to the parity  $f_n(-x)=(-1)^n f_n(x)$.
It can be seen that $\Delta_\alpha \sigma_{\rm R,L}(x)$   satisfies
\begin{eqnarray}
\left(\partial_x^2-\lambda(x)\right) \Delta_\alpha \sigma_{\rm R,L}(x)
=\frac{\partial \lambda(x)}{\partial \alpha} \sigma_{\rm R,L} (x)
-\int dy \frac{\partial \lambda(y)}{\partial \alpha} \sigma_{\rm R,L} (y) \sum_{n} f_n(y) f_n(x)=0\;, 
\end{eqnarray}
and it must vanishes at both boundaries.  In other words,  $\Delta_\alpha \sigma_{\rm R,L}(x)$   is a normalizable, zero-energy solution of 
a Schr\"odinger equation. With a positive definite potential $ \lambda(x)$, the only solution is 
\be   \Delta_\alpha \sigma_{\rm R,L}(x)\equiv 0\;.
\ee
It follows that  $\partial \sigma_{\rm R,L}/\partial \alpha $ can be expanded in $\{ f_n(x) \} $ as, 
\begin{eqnarray}
\frac{\partial \sigma_{\rm R,L}}{\partial \alpha}(x) = \sum_{n} b_n^{\rm R,L}  f_n(x)\;.
\end{eqnarray}
Here the summation with respect to the infinite modes is assumed to converge even around the boundary
since the $\alpha$ dependence does not affect UV modes. 
Therefore we find that their behavior for $x\sim -L/2$ is, 
\begin{eqnarray}
\frac{\partial \sigma_{\rm R, L}}{\partial \alpha}(x) \sim   B_{\rm R,L}
 \frac{\left(\frac L 2+x\right)}{\sqrt{ \frac{N}{2\pi} \log\frac{L_0}{\frac{L}2+x} } }\;,
\end{eqnarray}
 with constants $B_{\rm R,L}$ obtained by,  
 \begin{eqnarray}
B_{\rm R}= \frac{\partial W}{\partial \alpha}=\sum_{n}b_n^{\rm R} a^{\rm L}_n=\sum_{n} b_n^{\rm L} a_n^{\rm R}\;,\qquad 
B_{\rm L}=\sum_{n}b_n^{\rm L} a^{\rm L}_n=\sum_{n}b_n^{\rm R} a^{\rm R}_n\;,
\end{eqnarray}
where  $\{ a_n^{\rm L}\}$ and $\{a_n^{\rm R}\}$ are  defined as
\begin{eqnarray}
f_n(x)\sim a_n^{\rm L} 
 \frac{\left(\frac L 2+x\right)}{\sqrt{ \frac{N}{2\pi} \log\frac{L_0}{\frac{L}2+x} }},\quad a_n^{\rm R}= (-1)^P a_n^{\rm L}\; .
\end{eqnarray}

Using these properties, we find that  
\begin{eqnarray}
\lim_{x \to \pm \frac{L}2} \left(
\frac{\partial \sigma_{\rm L}(x)}{\partial \alpha} \sigma_{\rm L}'(x)+\frac{\partial \sigma_{\rm R}(x)}{\partial \alpha} \sigma_{\rm R}'(x)
\right)=
\lim_{x \to \pm \frac{L}2}  {\cal O} \left( \frac1{\log (\frac{L}2\mp x)}\right)=0\;,\nn
\lim_{x \to \pm \frac{L}2} \left(
\frac{\partial \sigma_{\rm L}(x)}{\partial \alpha} \sigma_{\rm R}'(x)+\frac{\partial \sigma_{\rm R}(x)}{\partial \alpha} \sigma_{\rm L}'(x)
\right)=
\lim_{x \to \pm \frac{L}2}  {\cal O} \left( \frac1{\log (\frac{L}2\mp x)}\right)=0\;.
\end{eqnarray}

\section{Numerical accuracy}\label{app:numacc}

In this section, we will try to quantify the numerical accuracy of the
solutions presented in this paper.
The handle we have from the relaxation method is given by the quantity
\be
\varepsilon_{\rm numerical} \equiv 
  \int_{-\frac{L}{2}}^{\frac{L}{2}} dx
  \left|
  \frac{N}{2}\sum_{n=1}^{n_{\rm max}} \frac{f_n(x)^2}{\omega_n}
  + \sigma_1(x)^2 + \sigma_2(x)^2 - r_{n_{\rm max}}\right|,
\ee
which is the numerically evaluated gap equation.
In the numerical calculations presented in this paper, we have used
$\varepsilon_{\rm numerical}=10^{-4}$.
If we here denote by $\langle x\rangle$ the numerical accuracy, it is
easy to see that 
\be
\left\langle 
\int_{-\frac{L}{2}}^{\frac{L}{2}} dx\; 
\left[\sigma_1(x)^2 + \sigma_2(x)^2\right]
\right\rangle
< \varepsilon_{\rm numerical} = 10^{-4}\,.
\ee
The numerical lattice we used has {\tt LEN} $=10^4$ lattice points;
thus locally on \emph{average}, we have
\be
\left|\langle\sigma_a(x)\rangle\right|
< \sqrt{\frac{\varepsilon_{\rm numerical}}{\texttt{LEN}}} = 10^{-4}\,.
\ee
Notice, however, the combination $\sigma_1^2+\sigma_2^2$ used in the
gap equation has the local error, on average,
\be
\left\langle\sigma_1(x)^2+\sigma_2(x)^2\right\rangle
< \frac{\varepsilon_{\rm numerical}}{\texttt{LEN}} = 10^{-8}\, ,
\ee
which is very small.

Using now the fictitious time-flow equation
\eqref{eq:lambdaevolution}, we can estimate the error of $\lambda$ as 
\be
\left|\int_{-\frac{L}{2}}^{\frac{L}{2}}dx\; \Delta\lambda(x)\right|
  < h_\tau \varepsilon_{\rm numerical}
  < \varepsilon_{\rm numerical}\,,
\ee
where we have defined $\Delta\lambda\equiv\lambda_{t+1}-\lambda_t$
and used that $h_\tau<1$.
Thus, we have also for $\lambda$, that the numerical precision on
\emph{average}, locally, is
\be
\langle\lambda(x)\rangle
< \frac{\varepsilon_{\rm numerical}}{\texttt{LEN}} = 10^{-8}\,.
\ee

Using error propagation, we can estimate the precision of the
Wronskian as
\be
\langle W\rangle <
\sqrt{\left(\frac{\varepsilon_{\rm numerical}}{\texttt{LEN}}\right)^2
  +\left(h_x^4\right)^2}
= \sqrt{\left(\frac{\varepsilon_{\rm numerical}}{\texttt{LEN}}\right)^2
  +\left(\frac{L}{\texttt{LEN}}\right)^8}
\simeq \frac{\varepsilon_{\rm numerical}}{\texttt{LEN}}
= 10^{-8}\,.
\ee
Indeed, in Fig.~\ref{fig:wronskian} it is not possible to see any
numerical error at the $10^{-6}$ level.

Let us now estimate the precision of the $\alpha$ derivative of the
energy using Eq.~\eqref{check}; starting with the first parenthesis,
we have
\be
\sqrt{\varepsilon_{\rm numerical}^2+\left(h_\alpha^4\right)^2}
=\sqrt{\varepsilon_{\rm numerical}^2+\left(\frac{\pi}{16}\right)^8}
\simeq 1.5\times 10^{-3}\,,
\ee
where we have neglected the error of the spatial derivative.
Ignoring factors of order one, we can see that the second parenthesis
of Eq.~\eqref{check} yields the same order of magnitude.
Hence,
\be
\left\langle\frac{\de E}{\de\alpha}\right\rangle
< \sqrt{\varepsilon_{\rm numerical}^2+\left(\frac{\pi}{16}\right)^8}
\simeq 1.5\times 10^{-3}\,.
\ee

It is a bit harder to estimate the precision of the alternative
formula for the $\alpha$ derivative of the energy,
Eq.~\eqref{checkBis}. 
Let us consider the error estimate of each term of the energy density
in turn. 
The first term (in the brackets) in Eq.~\eqref{eq:calE0def} can be
estimated as the error of $\sum_n\omega_n$ due to the normalization of
the eigenmodes, which we can write as
\be
\sqrt{n_{\rm max}}\frac{\varepsilon_{\rm numerical}}{\texttt{LEN}}
\ee
Ignoring the spatial derivatives (as they are quite precise, see
above), all the other terms in Eq.~\eqref{eq:calE0def} as well as
$\lambda$ itself have an error level of
$\varepsilon_{\rm numerical}$. 
Since, for $n_{\rm max}=350$ and {\tt LEN} $=10^4$, the largest error
is $\varepsilon_{\rm numerical}$, we get the net precision of the
$\alpha$ derivative using Eq.~\eqref{checkBis} is the same as using
Eq.~\eqref{check}.
In practice, however, the numerical error of using
Eq.~\eqref{checkBis} is a bit larger than of Eq.~\eqref{check}; the
difference is an order-one factor, which we did not evaluate.

\section{The solutions of the gap equation for $L=1,8$}\label{sec:L1and8}

\begin{figure}[!ht]
\begin{center}
\mbox{\subfloat{\includegraphics[width=0.49\linewidth]{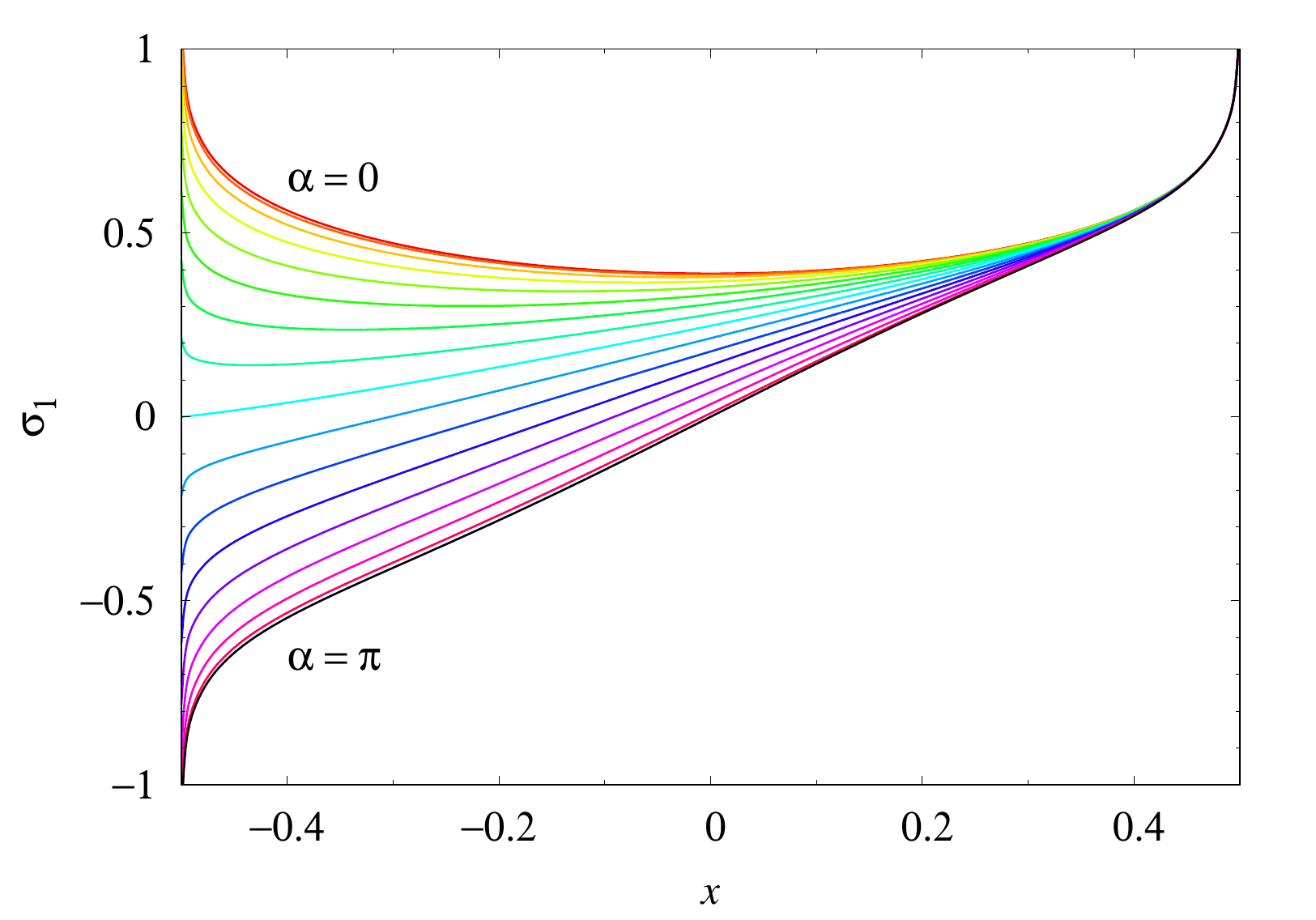}}
\subfloat{\includegraphics[width=0.49\linewidth]{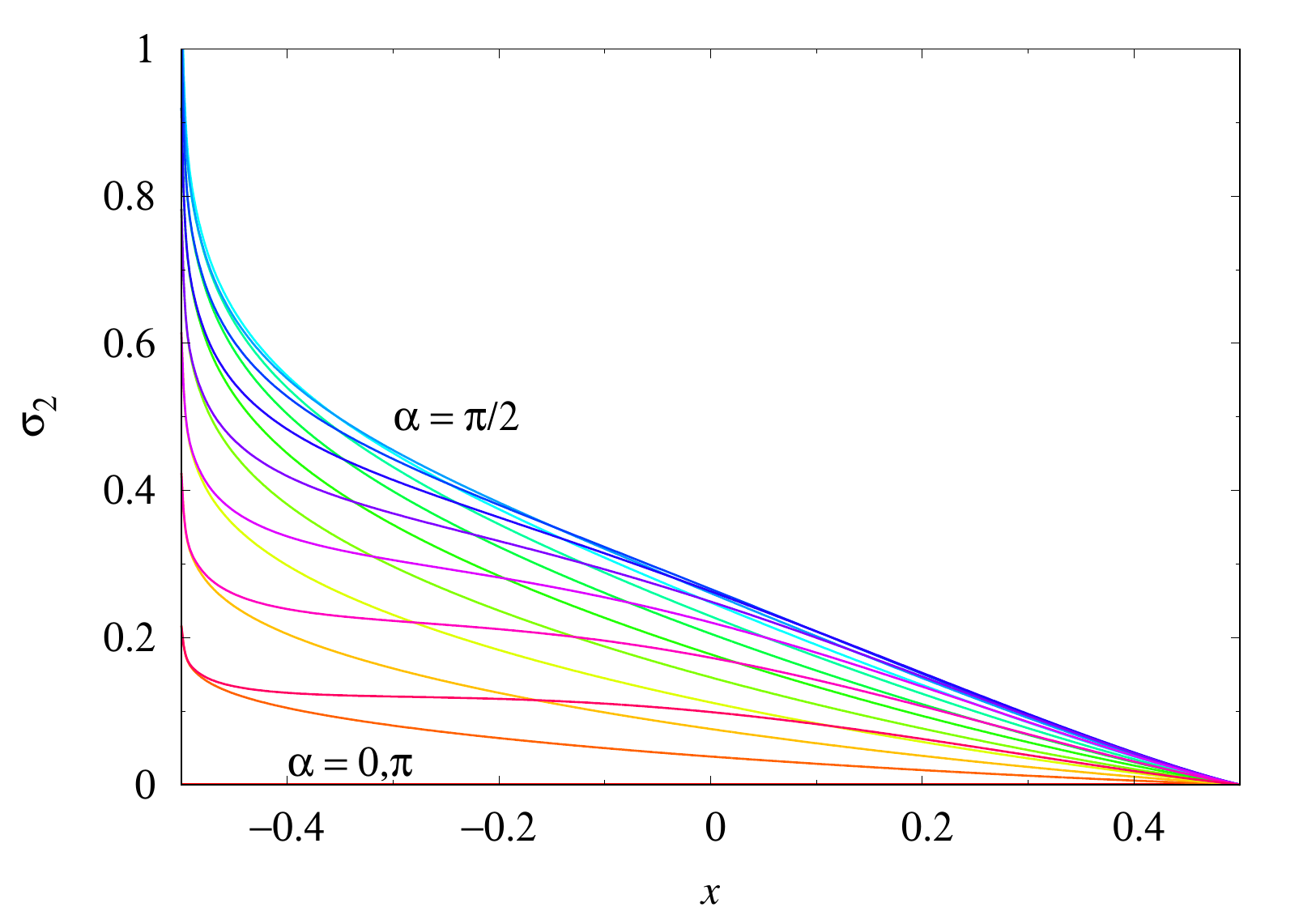}}}
\caption{\small  The same as Fig.~\ref{fig:sigmas} but for $L=1$.
}
\label{fig:sigmasBis}
\end{center}
\end{figure}

\begin{figure}[!ht]
\begin{center}
\mbox{\subfloat{\includegraphics[width=0.49\linewidth]{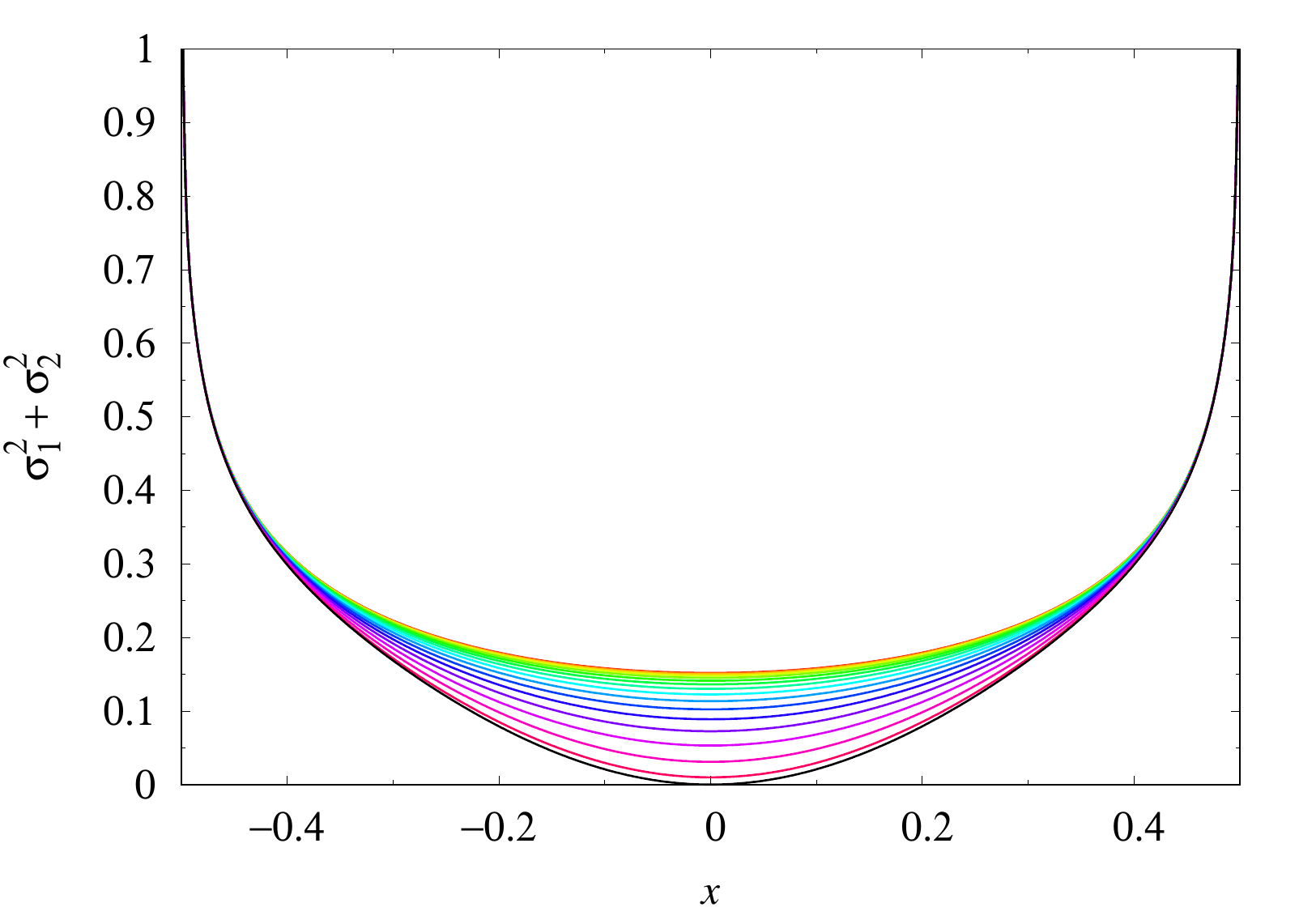}}
\subfloat{\includegraphics[width=0.49\linewidth]{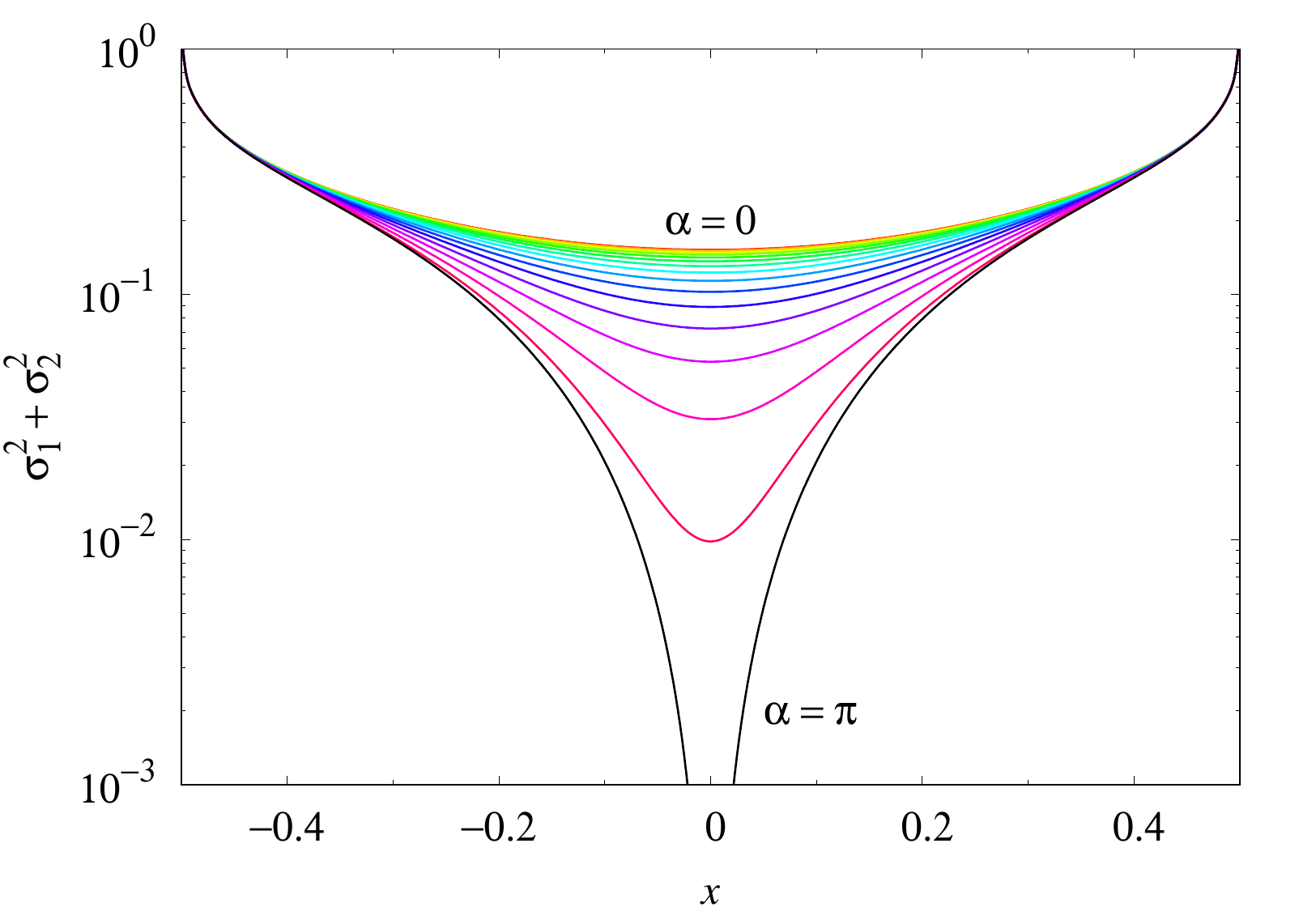}}}
\caption{\small The same as Fig.~\ref{fig:ssum} but for $L=1$.
}
\label{fig:ssumBis}
\end{center}
\end{figure}

\begin{figure}
\begin{center}
\includegraphics[width=3.5in]{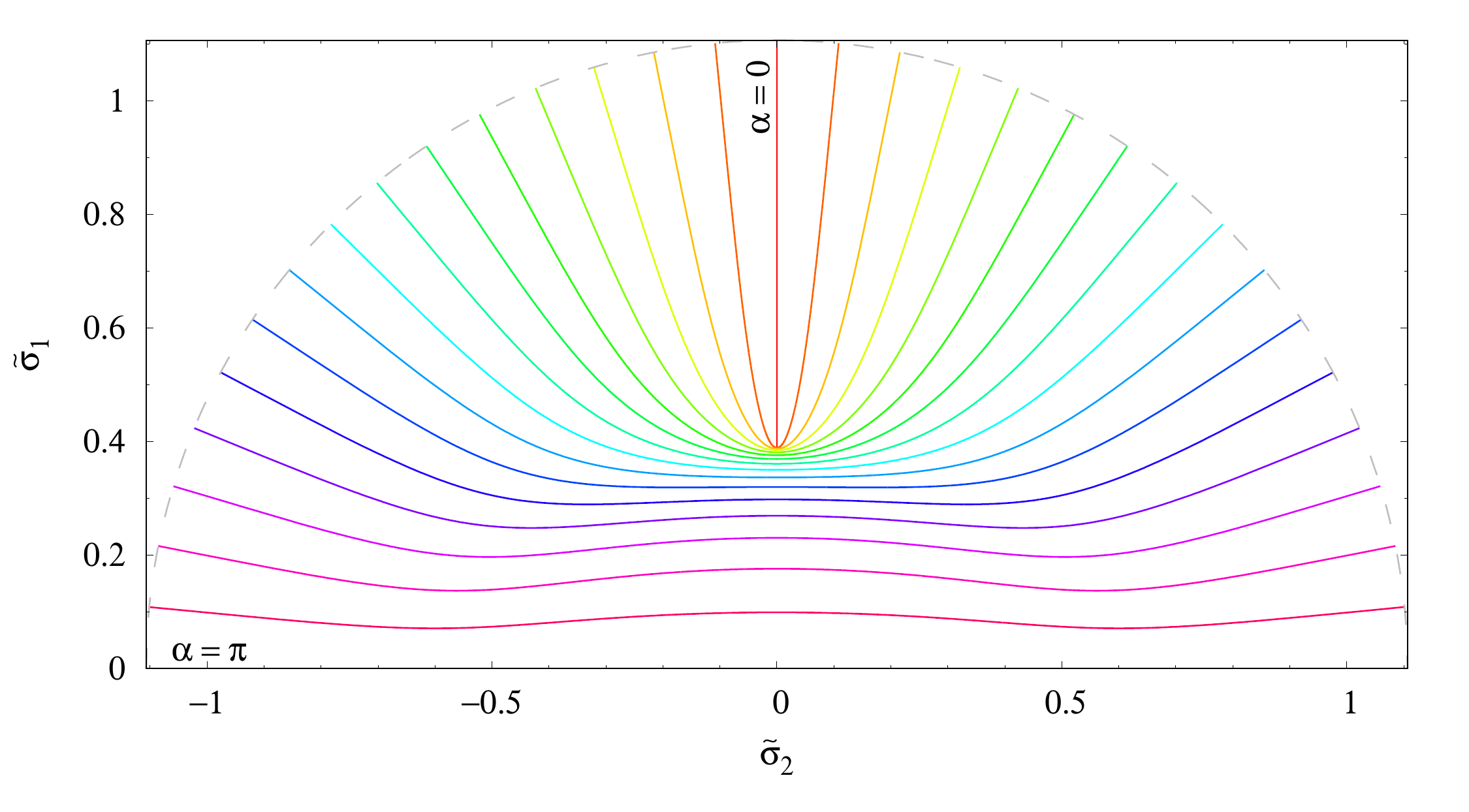}
\caption{\small  The same as Fig.~\ref{fig:sigmatilde} but for $L=1$.
}
\label{fig:sigmatildeBis}
\end{center}
\end{figure}

\begin{figure}[!ht]
\begin{center}
\mbox{\subfloat{\includegraphics[width=0.49\linewidth]{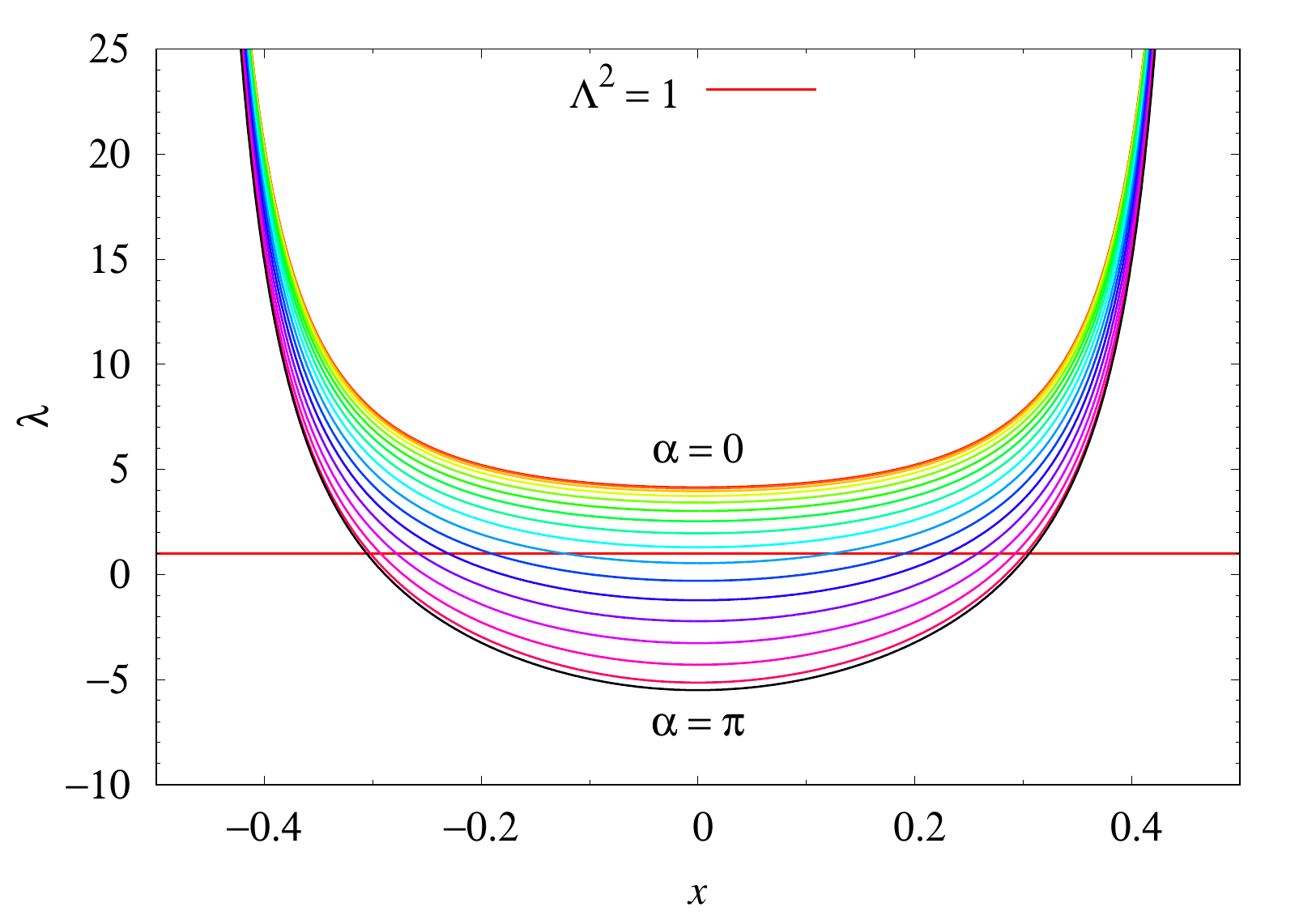}}
\subfloat{\includegraphics[width=0.44\linewidth]{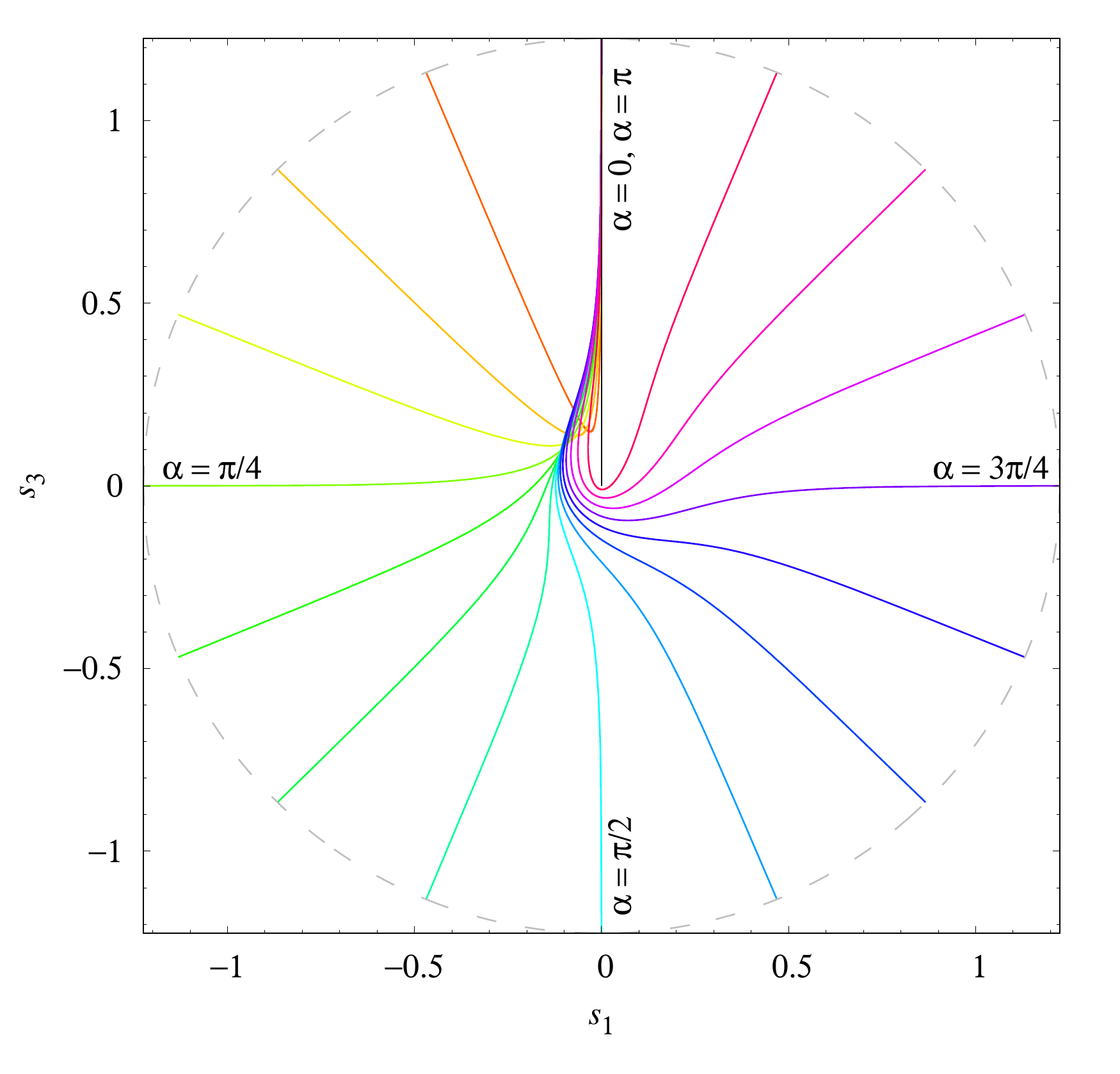}}}
\caption{\small  The same as Fig.~\ref{fig:lambda} (left) and
  Fig.~\ref{fig:O3} (right) but for $L=1$.
}
\label{fig:lambdaBis}
\end{center}
\end{figure}

\begin{figure}[!ht]
\begin{center}
\mbox{\subfloat{\includegraphics[width=0.49\linewidth]{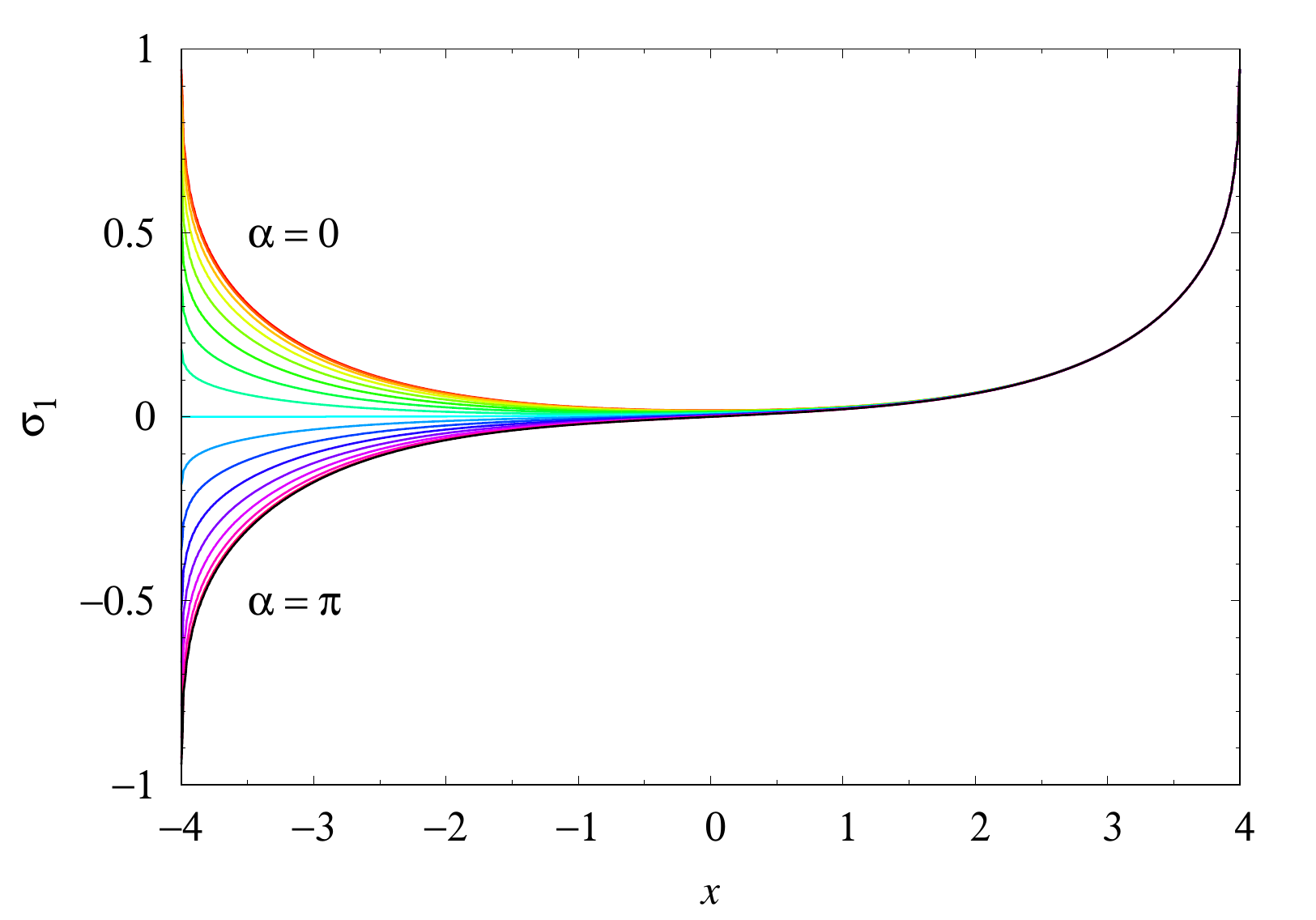}}
\subfloat{\includegraphics[width=0.49\linewidth]{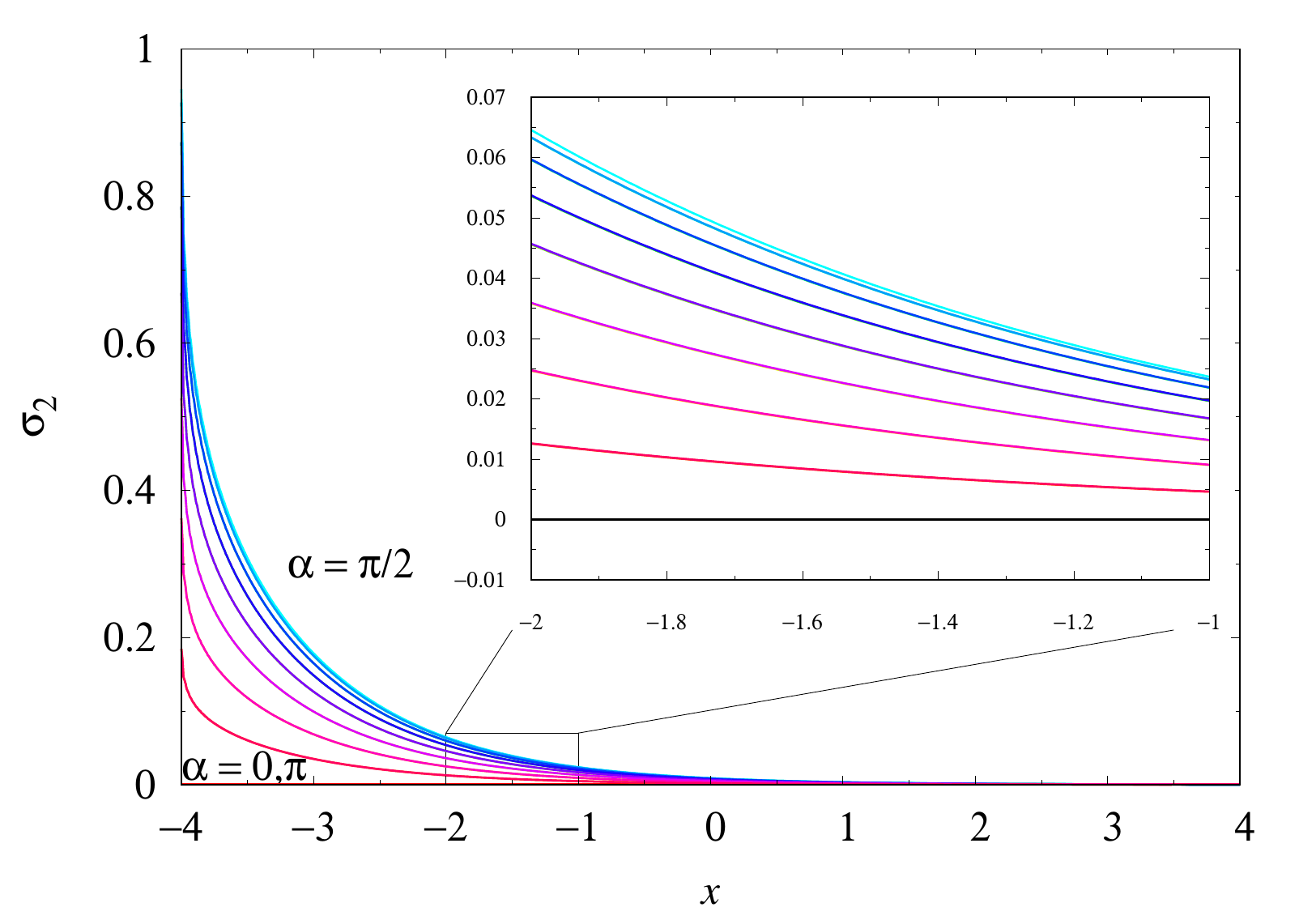}}}
\caption{\small  The same as Fig.~\ref{fig:sigmas} but for $L=8$.
}
\label{fig:sigmasBisBis}
\end{center}
\end{figure}

\begin{figure}[!ht]
\begin{center}
\mbox{\subfloat{\includegraphics[width=0.49\linewidth]{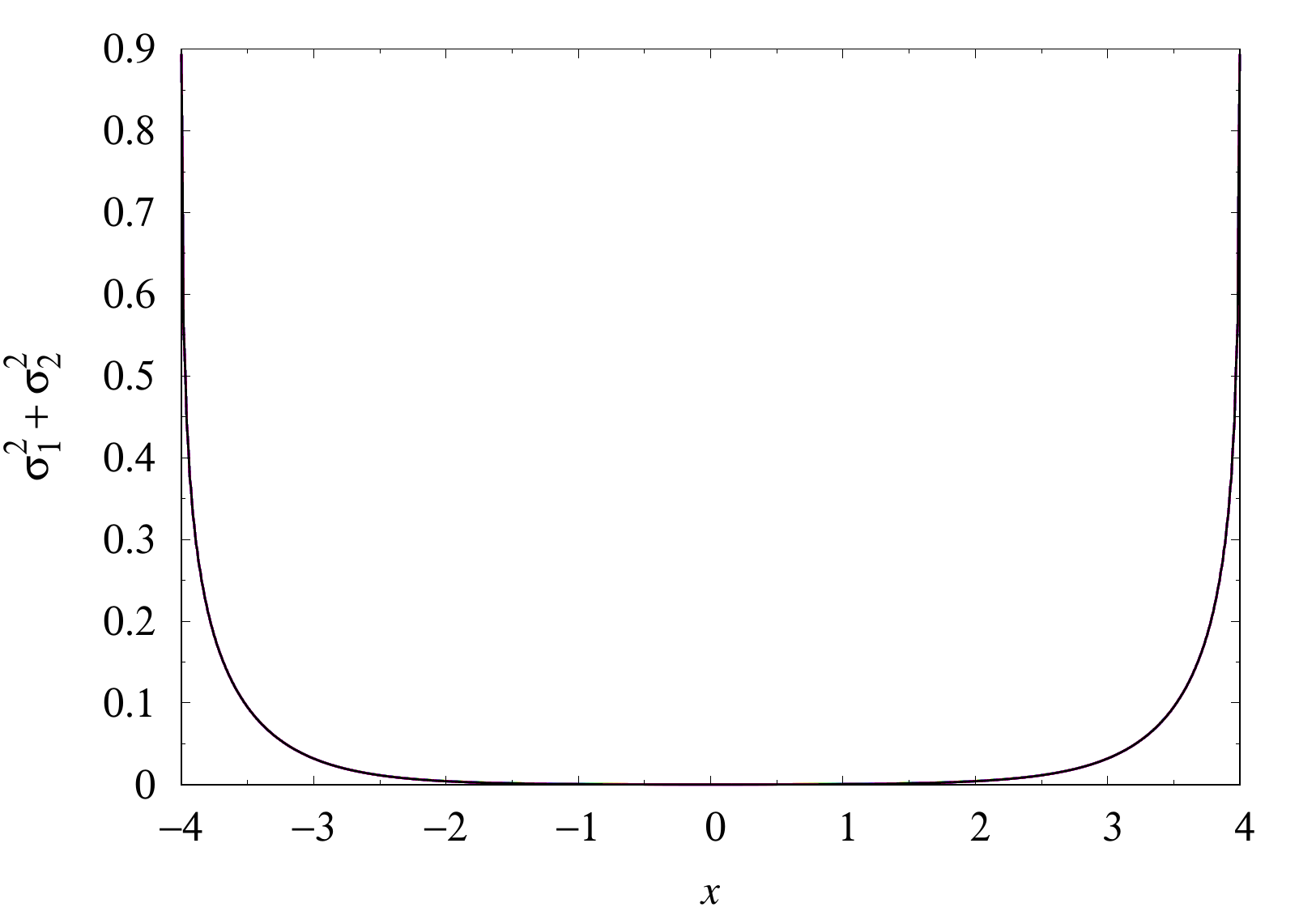}}
\subfloat{\includegraphics[width=0.49\linewidth]{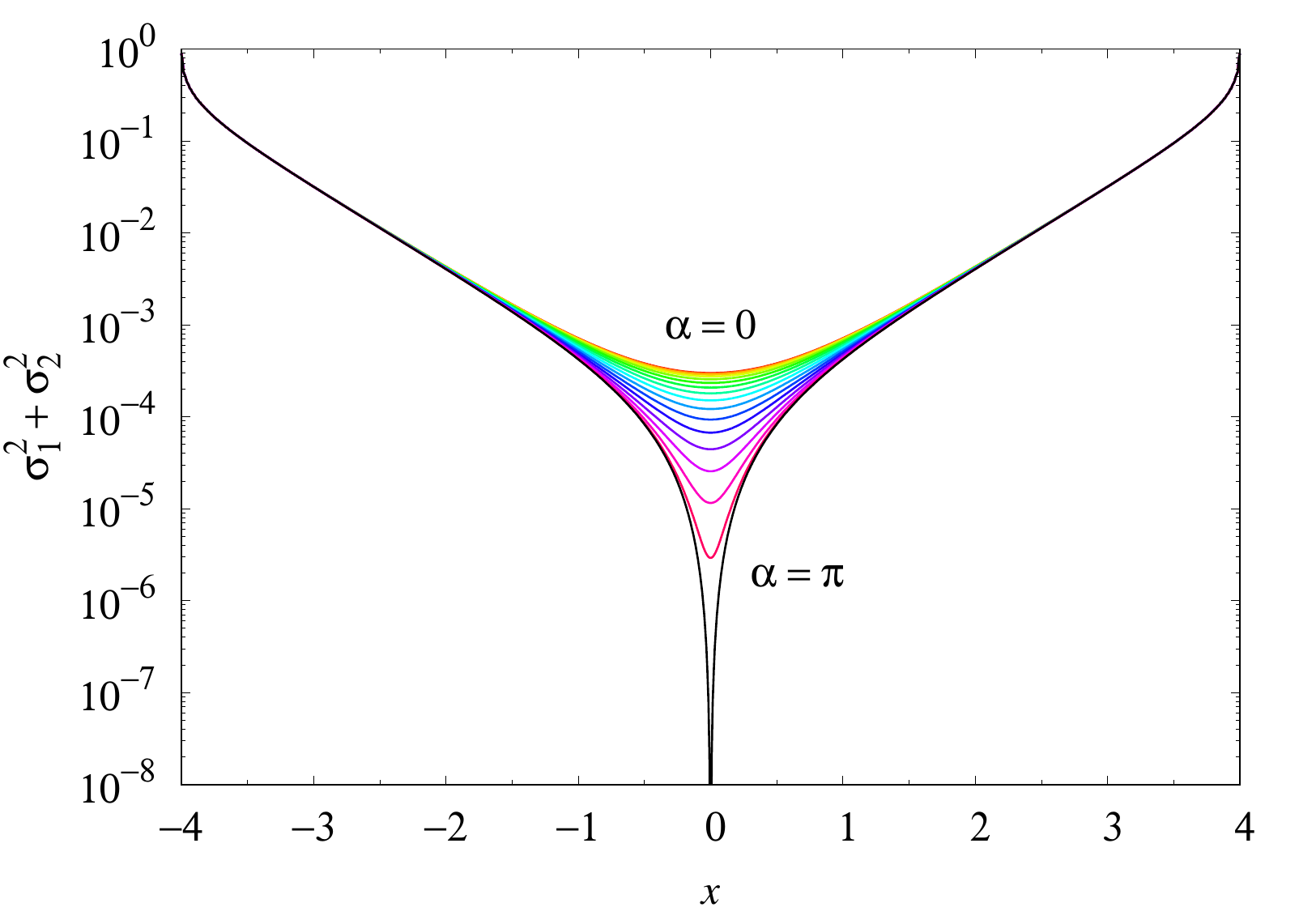}}}
\caption{\small The same as Fig.~\ref{fig:ssum} but for $L=8$.
}
\label{fig:ssumBisBis}
\end{center}
\end{figure}

\begin{figure}
\begin{center}
\includegraphics[width=3.5in]{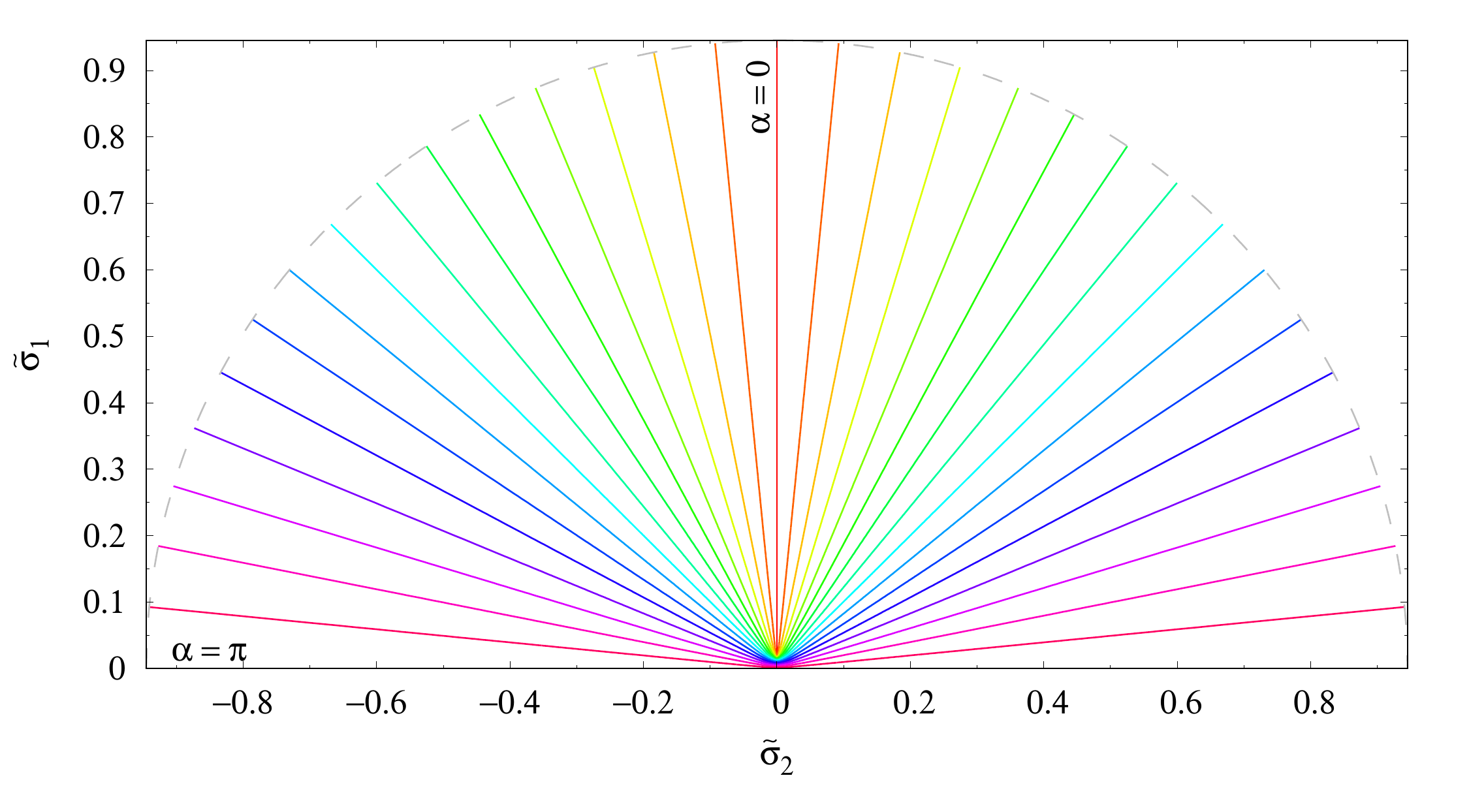}
\caption{\small  The same as Fig.~\ref{fig:sigmatilde} but for $L=8$.
}
\label{fig:sigmatildeBisBis}
\end{center}
\end{figure}

\begin{figure}[!ht]
\begin{center}
\mbox{\subfloat{\includegraphics[width=0.49\linewidth]{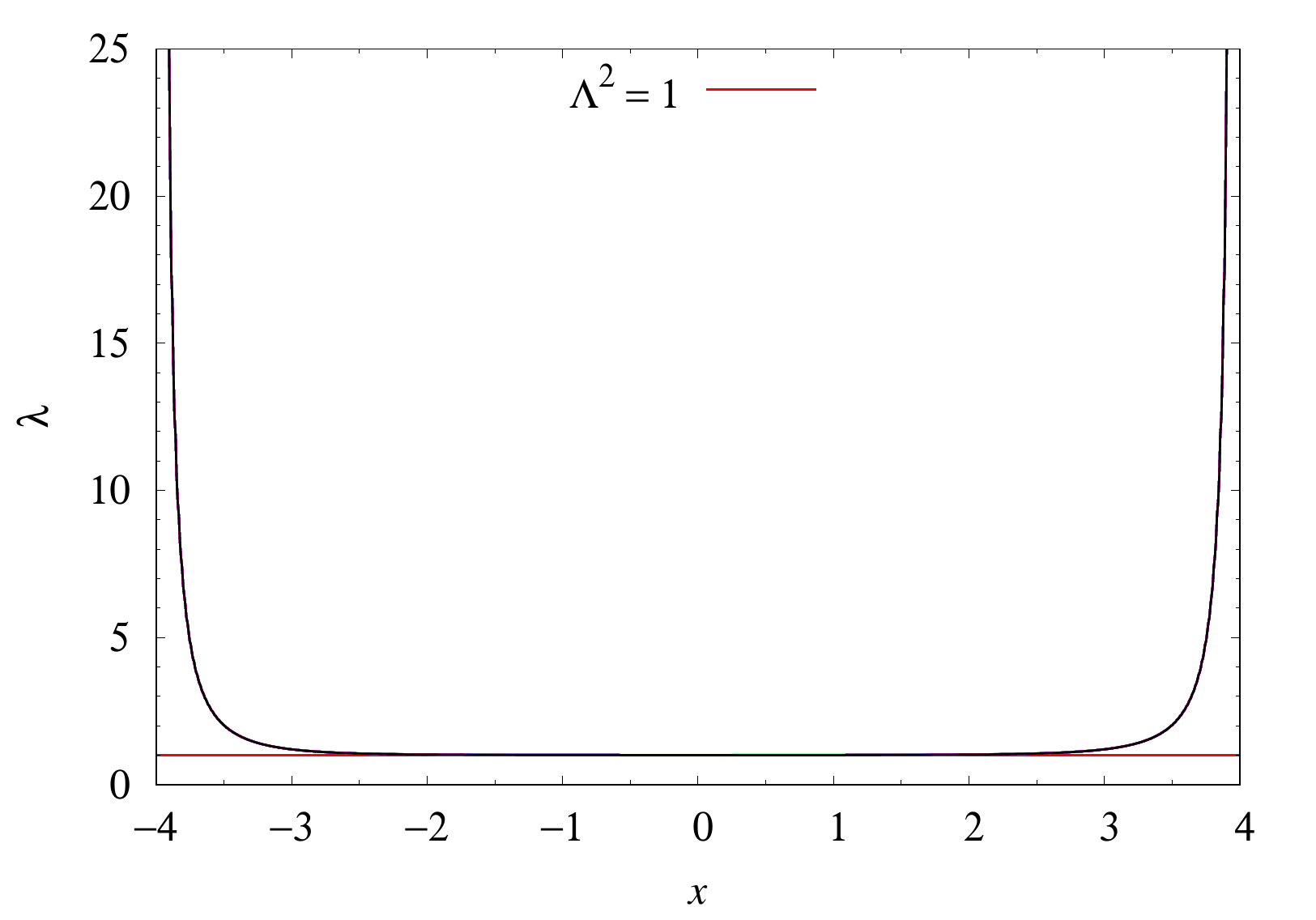}}
\subfloat{\includegraphics[width=0.49\linewidth]{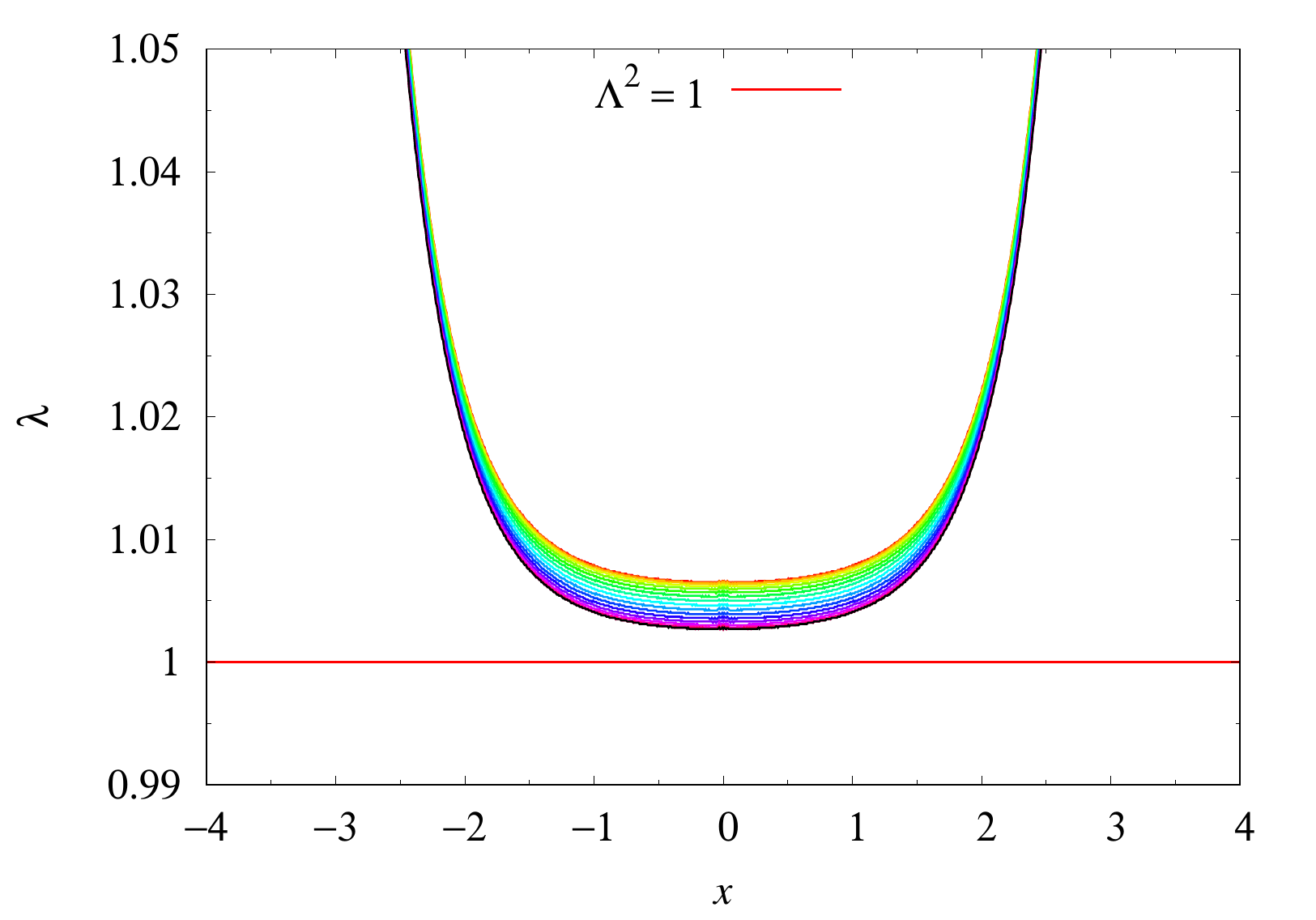}}}
\caption{\small  The same as Fig.~\ref{fig:lambda} but for $L=8$.
}
\label{fig:lambdaBisBis}
\end{center}
\end{figure}

\begin{figure}[!ht]
\begin{center}
\mbox{\subfloat{\includegraphics[width=0.44\linewidth]{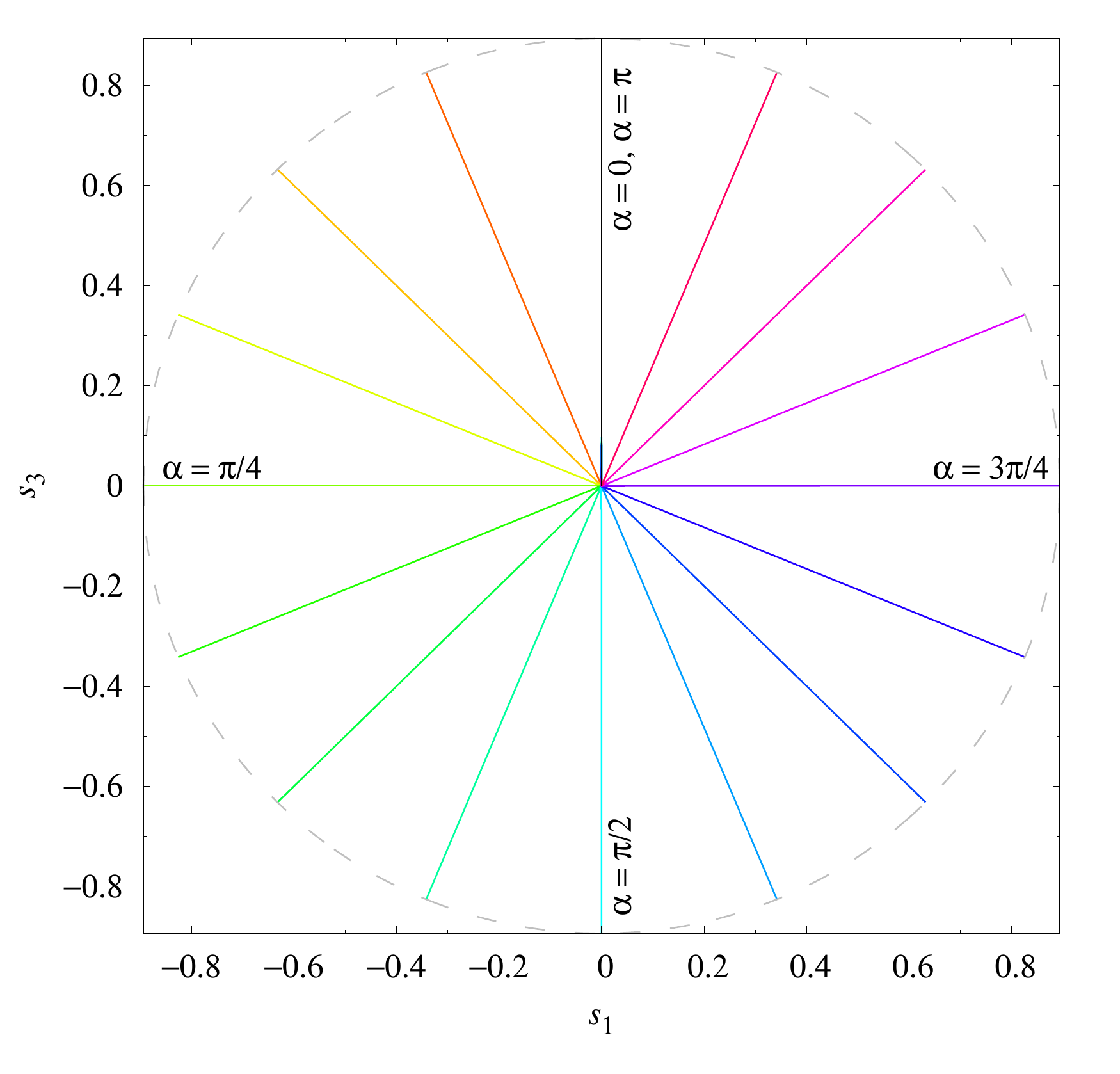}}
\subfloat{\includegraphics[width=0.44\linewidth]{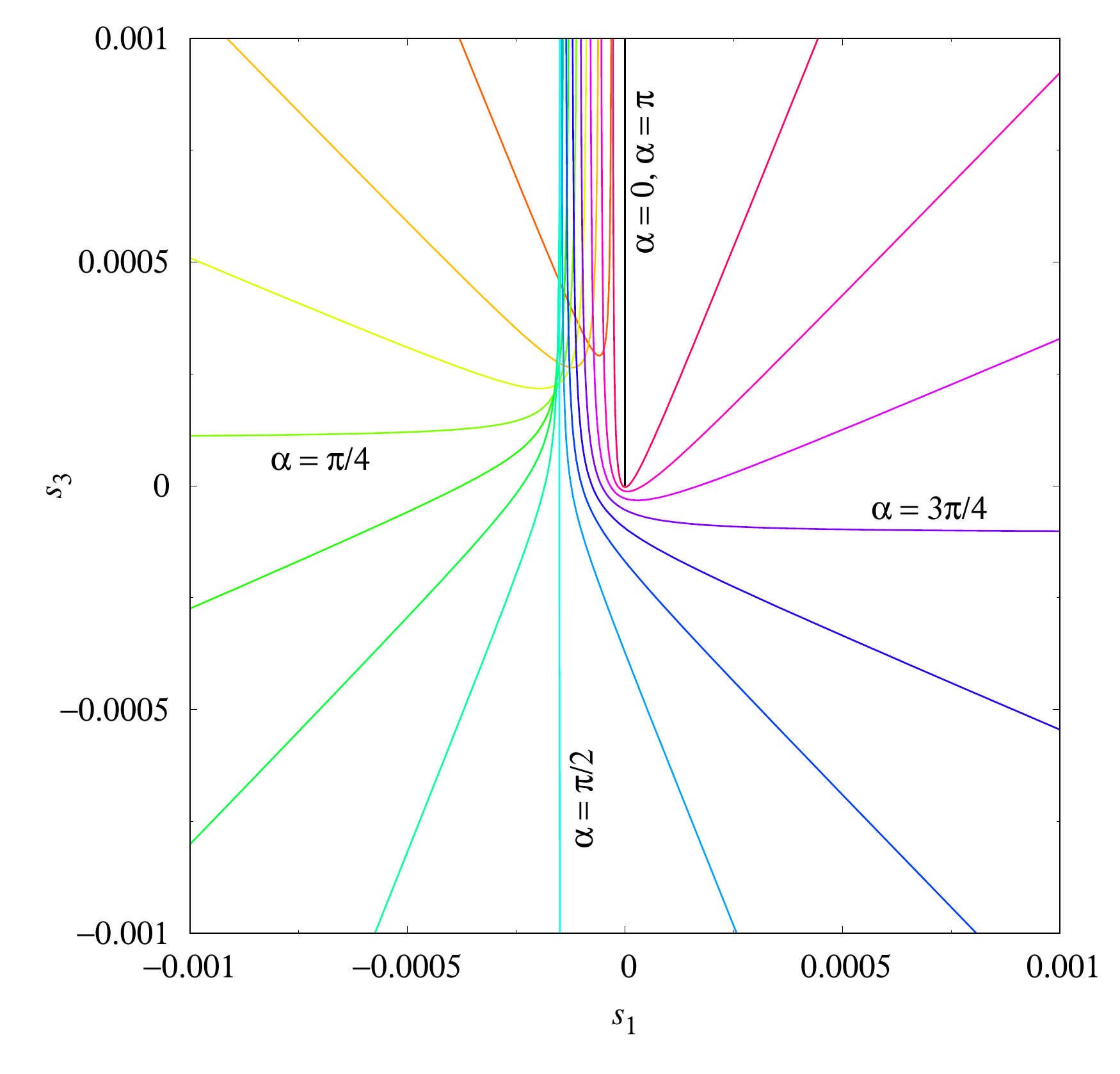}}}
\caption{\small  The same as Fig.~\ref{fig:O3} but for $L=8$.
}
\label{fig:O3BisBis}
\end{center}
\end{figure}

The results for $L=1$ and $L=8$ corresponding to
Figs.~\ref{fig:sigmas}-\ref{fig:O3} are shown in
Figs.~\ref{fig:sigmasBis}-\ref{fig:lambdaBis} and
Figs.~\ref{fig:sigmasBisBis}-\ref{fig:O3BisBis}.


\begin{thebibliography}{11}


\bibitem{DAdda:1978vbw}
  A.~D'Adda, M.~L\"uscher and P.~Di Vecchia,
  ``A 1/n Expandable Series of Nonlinear Sigma Models with Instantons,''
  Nucl.\ Phys.\ B {\bf 146} (1978) 63.
  doi:10.1016/0550-3213(78)90432-7

\bibitem{Witten:1978bc}
  E.~Witten,
  ``Instantons, the Quark Model, and the 1/n Expansion,''
  Nucl.\ Phys.\ B {\bf 149} (1979) 285.
  doi:10.1016/0550-3213(79)90243-8

\bibitem{Affleck:1984ar} 
  I.~Affleck,
  ``The Quantum Hall Effect, $\sigma$ Models at $\theta=\pi$ and Quantum Spin Chains,''
  Nucl.\ Phys.\ B {\bf 257} (1985) 397.
  doi:10.1016/0550-3213(85)90353-0

\bibitem{Sondhi:1993zz} 
  S.~L.~Sondhi, A.~Karlhede, S.~A.~Kivelson and E.~H.~Rezayi,
  ``Skyrmions and the crossover from the integer to fractional quantum Hall effect at small Zeeman energies,''
  Phys.\ Rev.\ B {\bf 47} (1993) 16419.
  doi:10.1103/PhysRevB.47.16419
  
\bibitem{Ezawa:1999}
  Z.~F.~Ezawa,
  ``Spin-Pseudospin Coherence and $CP^3$ Skyrmions in Bilayer Quantum
  Hall Ferromagnets,''
  Phys.\ Rev.\ Lett.\ {\bf 82} (1999) 3512.

\bibitem{Arovas:1999}
  D.~P.~Arovas, A.~Karlhede, and D.~Lilliehook,
  ``$SU(N)$ quantum Hall skyrmions,''
  Phys.\ Rev.\ B {\bf 59} (1999) 13147 .
  
\bibitem{Rajaraman:2002}
  R.~Rajaraman,
  ``$CP_N$ solitons in quantum Hall systems,''
  Eur.\ Phys.\ J.\ B {\bf 29} (2002) 157 .
  doi:10.1140/epjb/e2002-00277-7

\bibitem{BKO}
  S.~Bolognesi, K.~Konishi and K.~Ohashi,
  ``Large-$N$ $\mathbb{C}P^{N-1}$ sigma model on a finite interval,''
  JHEP {\bf 1610} (2016) 073
  doi:10.1007/JHEP10(2016)073
  [arXiv:1604.05630 [hep-th]].
  
\bibitem{BBGKO}
  A.~Betti, S.~Bolognesi, S.~B.~Gudnason, K.~Konishi and K.~Ohashi,
  ``Large-N CP(N-1) sigma model on a finite interval and the renormalized string energy,''
  arXiv:1708.08805 [hep-th], JHEP (2018), to appear.
  
\bibitem{MVComplex1}
  R.~Auzzi, S.~Bolognesi, J.~Evslin and K.~Konishi,
  ``NonAbelian monopoles and the vortices that confine them,''
  Nucl.\ Phys.\ B {\bf 686} (2004) 119
  doi:10.1016/j.nuclphysb.2004.03.003
  [hep-th/0312233].

\bibitem{MVComplex2}
  K.~Konishi, A.~Michelini and K.~Ohashi,
  ``Monopole-vortex complex in a theta vacuum,''
  Phys.\ Rev.\ D {\bf 82} (2010) 125028
  doi:10.1103/PhysRevD.82.125028
  [arXiv:1009.2042 [hep-th]].

\bibitem{MVComplex3}
  M.~Cipriani, D.~Dorigoni, S.~B.~Gudnason, K.~Konishi and A.~Michelini,
  ``Non-Abelian monopole-vortex complex,''
  Phys.\ Rev.\ D {\bf 84} (2011) 045024
  doi:10.1103/PhysRevD.84.045024
  [arXiv:1106.4214 [hep-th]].

\bibitem{MVComplex4}
  C.~Chatterjee and K.~Konishi,
  ``Monopole-vortex complex at large distances and nonAbelian duality,''
  JHEP {\bf 1409} (2014) 039
  doi:10.1007/JHEP09(2014)039
  [arXiv:1406.5639 [hep-th]].
  
\bibitem{Hanany:2003hp}
  A.~Hanany and D.~Tong,
  ``Vortices, instantons and branes,''
  JHEP {\bf 0307} (2003) 037
  doi:10.1088/1126-6708/2003/07/037
  [hep-th/0306150].

\bibitem{Auzzi:2003fs}
  R.~Auzzi, S.~Bolognesi, J.~Evslin, K.~Konishi and A.~Yung,
  ``NonAbelian superconductors: Vortices and confinement in N=2 SQCD,''
  Nucl.\ Phys.\ B {\bf 673} (2003) 187
  doi:10.1016/j.nuclphysb.2003.09.029
  [hep-th/0307287].

\bibitem{Shifman:2004dr}
  M.~Shifman and A.~Yung,
  ``NonAbelian string junctions as confined monopoles,''
  Phys.\ Rev.\ D {\bf 70} (2004) 045004
  doi:10.1103/PhysRevD.70.045004
  [hep-th/0403149].
  
\bibitem{Milekhin:2012ca}
  A.~Milekhin,
  ``CP(N-1) model on finite interval in the large N limit,''
  Phys.\ Rev.\ D {\bf 86} (2012) 105002
  doi:10.1103/PhysRevD.86.105002
  [arXiv:1207.0417 [hep-th]].
   
\bibitem{Monin:2015xwa}
  S.~Monin, M.~Shifman and A.~Yung,
  ``Non-Abelian String of a Finite Length,''
  Phys.\ Rev.\ D {\bf 92} (2015) no.2,  025011
  doi:10.1103/PhysRevD.92.025011
  [arXiv:1505.07797 [hep-th]].
  
\bibitem{Milekhin:2016fai}
  A.~Milekhin,
  ``CP(N) sigma model on a finite interval revisited,''
  Phys.\ Rev.\ D {\bf 95} (2017) no.8,  085021
  doi:10.1103/PhysRevD.95.085021
  [arXiv:1612.02075 [hep-th]].

\bibitem{Flachi:2017cdo}
  A.~Flachi, M.~Nitta, S.~Takada and R.~Yoshii, 
  ``Sign Flip in the Casimir Force for Interacting Fermion Systems,''
  Phys.\ Rev.\ Lett.\ {\bf 119} (2017) no.3, 031601
  doi:10.1103/PhysRevLett.119.031601
  [arXiv:1704.04918 [hep-th]].

\bibitem{Nitta:2017uog}
  M.~Nitta and R.~Yoshii,
  ``Self-Consistent Large-$N$ Analytical Solutions of Inhomogneous Condensates in Quantum ${\mathbb C}P^{N-1}$ Model,''
  arXiv:1707.03207 [hep-th].
  
\bibitem{NittaYoshiiNew}
  A.~Flachi, M.~Nitta, S.~Takada and R.~Yoshii,
  ``Casimir Force for the ${\mathbb C}P^{N-1}$ Model,''
  arXiv:1708.08807 [hep-th].
  
\bibitem{Pavshinkin}
  D.~Pavshinkin,
  ``Grassmannian sigma model on a finite interval,''
  arXiv:1708.06399 [hep-th].
  
\bibitem{Nitta:2018lnn} 
  M.~Nitta and R.~Yoshii,
  ``Self-consistent Analytic Solutions in Twisted $\mathbb{C}P^{N-1}$ Model in the Large-$N$ Limit,''
  arXiv:1801.09861 [hep-th].
 


\end{thebibliography}
\end{document}